\documentclass[sigconf]{acmart}
\usepackage{comment}
\usepackage{float}
\usepackage{csquotes}
\usepackage{tabularx}
\usepackage{xcolor}
\usepackage{graphicx}
\usepackage{subcaption}
\usepackage{framed}

\newcommand{\edited}[1]{#1}

\newcommand{\leftcell}[2][l]{%
  \begin{tabular}[#1]{@{}l@{}}#2\end{tabular}}
  
\newcommand{\circledigit}[1]{\textbf{\normalsize{\textsf{\textcircled{\footnotesize{#1}}}}}}

\newcommand{\eg}{\textit{e.g.}}
\newcommand{\ie}{\textit{i.e.}}

\newcommand{\etal}{\textit{et al.}}
\newcommand{\system}{PosterMate} 

\newlength\myindent
\AtBeginDocument{%
  \providecommand\BibTeX{{%
    \normalfont B\kern-0.5em{\scshape i\kern-0.25em b}\kern-0.8em\TeX}}}

\makeatletter
\def\@ACM@copyright@check@cc{}
\makeatother

\copyrightyear{2025}
\acmYear{2025}
\setcopyright{cc}
\setcctype{by}
\acmConference[UIST '25]{The 38th Annual ACM Symposium on User Interface Software and Technology}{September 28-October 1, 2025}{Busan, Republic of Korea}
\acmBooktitle{The 38th Annual ACM Symposium on User Interface Software and Technology (UIST '25), September 28-October 1, 2025, Busan, Republic of Korea}\acmDOI{10.1145/3746059.3747769}
\acmISBN{979-8-4007-2037-6/2025/09}

\begin{document}
\newcommand\todoit[1]{{\color{red}\{TODO: \textit{#1}\}}}
\newcommand\todocite{{\color{red}{CITE}}}

\definecolor{lightblue}{RGB}{212, 235, 255}
\definecolor{orange}{RGB}{255, 105, 0}
\definecolor{lightgreen}{RGB}{177, 231, 171}
\definecolor{lightyellow}{RGB}{255, 255, 148}

\newcolumntype{P}[1]{>{\centering\arraybackslash}p{#1}}
\newcolumntype{L}[1]{>{\raggedright\let\newline\\\arraybackslash\hspace{0pt}}m{#1}}
\newcolumntype{R}[1]{>{\raggedleft\arraybackslash}p{#1}}
\newcommand\tworows[1]{\multirow{2}{*}{\shortstack[l]{#1}}}
\newcommand\tworowsc[1]{\multirow{2}{*}{\shortstack[c]{#1}}}
\newcommand\threerows[1]{\multirow{3}{*}{\shortstack[l]{#1}}}

\title[\system{}]{\system{}: Audience-driven Collaborative Persona Agents for Poster Design}

\author{Donghoon Shin}
\orcid{0000-0001-9689-7841}
\authornote{Work done during an internship at Adobe Research.}
\affiliation{%
 \institution{University of Washington}
 \city{Seattle}
 \state{WA}
 \country{USA}}
 \email{dhoon@uw.edu}

\author{Daniel Lee}
\orcid{0009-0005-3061-1827}
\affiliation{%
  \institution{Adobe Inc.}
  \city{San Jose}
  \state{CA}
  \country{USA}}
  \email{dlee1@adobe.com}

\author{Gary Hsieh}
\orcid{0000-0002-9460-2568}
\affiliation{%
 \institution{University of Washington}
 \city{Seattle}
 \state{WA}
 \country{USA}}
 \email{garyhs@uw.edu}
 
\author{Gromit Yeuk-Yin Chan}
\orcid{0000-0003-1356-4406}
\affiliation{%
  \institution{Adobe Research}
  \city{San Jose}
  \state{CA}
  \country{USA}}
  \email{ychan@adobe.com}

\renewcommand{\shortauthors}{Shin, et al.}

\begin{abstract}
\sloppy
Poster designing can benefit from synchronous feedback from target audiences. However, gathering audiences with diverse perspectives and reconciling them on design edits can be challenging. Recent generative AI models present opportunities to simulate human-like interactions, but it is unclear how they may be used for feedback processes in design. We introduce \system{}, a poster design assistant that facilitates collaboration by creating audience-driven persona agents constructed from marketing documents. \system{} gathers feedback from each persona agent regarding poster components, and stimulates discussion with the help of a moderator to reach a conclusion. These agreed-upon edits can then be directly integrated into the poster design. Through our user study ($N=12$), we identified the potential of \system{} to capture overlooked viewpoints, while serving as an effective prototyping tool. Additionally, our controlled online evaluation ($N=100$) revealed that the feedback from an individual persona agent is appropriate given its persona identity, and the discussion effectively synthesizes the different persona agents' perspectives.
\end{abstract}



\begin{teaserfigure}
  \includegraphics[width=\textwidth]{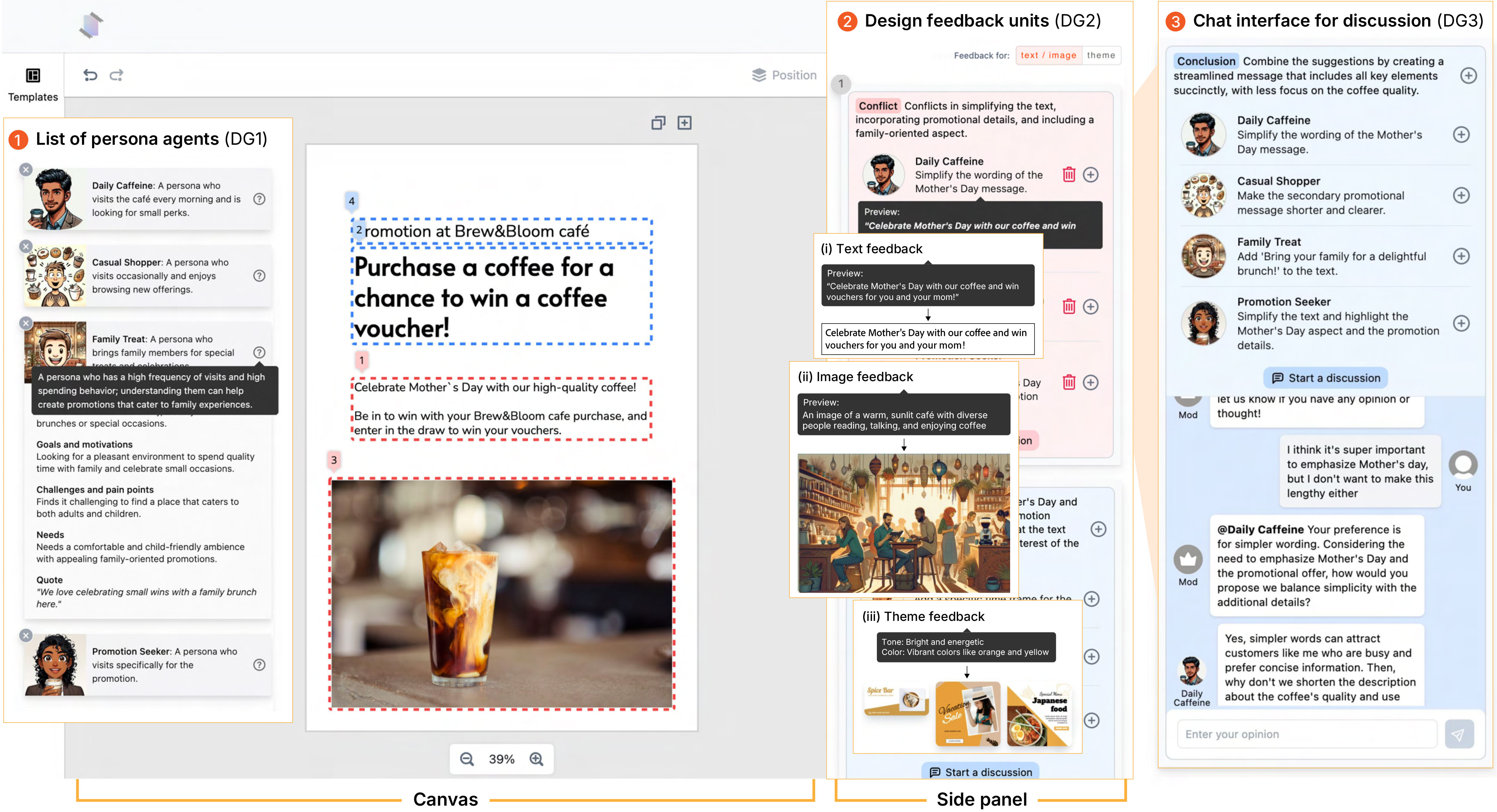}
  \caption{An overview of \system{}. Consisting of a canvas (left) and a side panel (right), \system{} assists a user in \circledigit{1} creating a set of persona agents based on the marketing brief, each of which \circledigit{2} provides a set of design feedback on the (i) text, (ii) image, and (iii) theme for the poster draft, along with its preview presented on hover. \circledigit{3} The user can also discuss with the persona agents on a design conflict to reach a conclusion.}
  \label{fig:keyscreen}
  \Description{Figure 1 is an overview of the PosterMate user interface, which is split into a left canvas area and a right side panel. The canvas shows a poster design for a café. The side panel is divided into three sections, indicated by numbered labels. (1) List of persona agents shows four AI personas with descriptions. (2) Design feedback units shows conflicting AI-generated feedback for the poster's text, image, and theme, with previews of the suggestions available on hover. (3) Chat interface for discussion displays a conversation where a moderator agent and persona agents discuss design conflicts to reach a resolution.}
\end{teaserfigure}

\maketitle

\section{Introduction}

An advertisement poster is a visual intended to convey a brand message, promote a product or event, and capture the attention of a target audience~\cite{guo2021vinci}, by delivering content aligned with their needs and traits~\cite{amplifyoutdoorbenefits2024}. To achieve this, it is valuable for designers to gather feedback from diverse audiences within the marketing target directly to make the design appeal to a broad range of intended viewers~\cite{forlano2018critique, mccormack2020design} through collaborative feedback processes~\cite{kang2018paragon, greenberg2015critiki}. Prior research has explored the use of human computational methods to engage potential feedback givers in the design process~\cite{xu2014voyant}, and keep these individuals on retainer for synchronous feedback while ensuring consistency whenever design-related questions arise~\cite{losev2020distributed, bernstein2011crowds}. However, these approaches can be costly and time-consuming---especially when recruiting specific target audiences as opposed to general crowds---and can undermine agile explorations.

But what if we construct target audiences as agents and have them provide feedback during the design process? Recent advancements in generative AI, particularly in simulating human behavior through agents~\cite{park2023generative, park2022social, qian2024chatdev}, present unique opportunities to address these challenges. These works offer opportunities to harness large language models (LLMs) for generating insights driven by human behaviors through agent-driven processes and simulating social interactions~\cite{park2022social}. With the benefits of interacting with the target audiences in uncovering active and latent needs (\eg{}, audience testing), we hypothesized that constructing agents that reflect target audiences and having them in the design feedback process would help designers lower the barrier towards capturing the needs of the target audiences. Yet, it remains unclear how these simulations could be operationalized and effectively applied to real-world design scenarios like poster designing. Also, no workflows have been proposed to facilitate meaningful, design-oriented discussions among such agents.

\sloppy
In this work, we propose a novel approach to incorporating simulations of target audiences in the poster design feedback process. Based on our preliminary exploration of design space, we introduce \system{}, an audience-driven persona agent system for advertisement poster design. Leveraging generative AI (\ie{}, LLMs, text-to-image models), \system{} creates a diverse set of persona agents based on existing observations/research data on the target audience (\ie{}, marketing briefs). These persona agents produce design feedback on individual poster components that align with their specific needs, which can engage in iterative discussions with the user to resolve their conflicts on design feedback and reach conclusions. Additionally, both feedback from individual agents and the conclusions from these discussions could be applied to the actual design outcome in real-time, based on the designer's discretion.

\sloppy
To evaluate the effectiveness of our audience-driven agents, we conducted an end-to-end user study and a controlled evaluation. In our user study ($N=12$), participants found \system{} useful and appreciated how persona-driven discussions helped them identify previously overlooked perspectives. Furthermore, in our online controlled evaluation ($N=100$), we found that feedback from an individual persona agent is perceived as appropriate given its persona identity, and the conclusion derived from the discussion effectively synthesizes the different audience personas' perspectives. Based on these findings, we discuss future implications for the use of persona agents in design feedback processes.

Our work contributes:
\begin{itemize}
    \item \system{}, a system that employs a novel approach of enabling design feedback and discussions through audience-driven persona agents to offer diverse perspectives and enrich advertisement poster design.
    \item Results from a user study and a controlled evaluation, demonstrating the potential of real-time collaborative design feedback processes involving persona agents.
    \item Key insights and design implications for employing audience-driven persona agents for effective design feedback and discussions.
\end{itemize}
\section{Related Work}

\subsection{Technology-mediated Design Feedback}

Design feedback is a core constituent of design, and extensive research in HCI has explored technology-aided supports to aid in these efforts. While these supports can vary widely in terms of actions and the degree of concurrency, prior research has often emphasized the importance of supporting diverse and synchronous feedback to improve the effectiveness, inclusivity, and sensemaking of the feedback~\cite{arias2000transcending, o2018charrette}.

The recognized advantages of design feedback have led many commercial design systems (\eg{}, Figma~\cite{figma} and Canva~\cite{canva}) to adopt features that support collaborative design feedback, such as anchoring comments and overlaying visual annotations to design components. Beyond these real-time online canvases, prior studies in HCI have attempted to incorporate modalities beyond visual canvas to support design feedback, including shared context~\cite{oehlberg2012showing}, hybrid physical-digital interactions~\cite{ishii1992clearboard, klemmer2002web}, and video conferencing~\cite{marlow2016beyond}.

A key assumption underlying feedback-driven design is the presence of sufficient individuals who can offer constructive feedback. However, identifying and engaging these individuals can be both time-consuming and challenging. To mitigate this issue, prior research has explored crowdsourcing as a strategy to collect diverse feedback when professional collaborators are unavailable~\cite{bernstein2011crowds, luther2014crowdcrit, xu2014voyant}. Yet, while crowds can supply a large quantity of feedback, its quality is often deemed lower than that from expert or peer designers~\cite{yuan2016almost}, prompting efforts to enhance the quality of crowd-generated feedback through interaction techniques (\eg{},~\cite{yuan2016almost, wu2013crowdcritter}). Nevertheless, crowdsourcing remains a costly and time-intensive approach---particularly when aiming to obtain feedback from specific user groups rather than a general audience. This poses a significant challenge in evaluating how well a design satisfies the needs of its intended audience from their own perspectives.

To address this gap, our work introduces computational agents as an alternative to human feedback providers in the design feedback process. Specifically, we propose constructing AI agents informed by target audience descriptions from marketing data to generate diverse feedback and participate in real-time discussions.

\subsection{AI-assisted Creativity Support Tools}

Previous research in HCI has explored various technological supports aimed at augmenting users' creativity and enhancing their creative work~\cite{frich2019mapping}. With recent advancements in AI, particularly in analyzing data and deriving insights~\cite{hwang2022too}, design recommendations have become increasingly tailored and multimodal. For instance, AI-powered tools have been shown to assist artists in generating new ideas~\cite{gonzalez2024collaborative, figoli2022ai}, refining existing concepts~\cite{zhong2024ai}, and providing inspiration for creative works~\cite{shin2024paper}.

Recent breakthroughs in generative AI, especially large language models (LLMs) and text-to-image (TTI) models, have opened up new opportunities for enhancing creativity with greater contextual relevance and personalized reasoning. For example, LLMs like GPT~\cite{achiam2023gpt} can generate coherent and contextually relevant text based on user prompts, and prior studies have demonstrated their potential to support writers in overcoming writer's block or exploring new narrative directions~\cite{lee2022coauthor, chung2022talebrush}. Similarly, text-to-image models (\eg{}, DALL-E 3~\cite{dalle3}, Adobe Firefly~\cite{firefly}) have been integrated into creative tools to help visual artists engage more deeply with their creative processes~\cite{ko2023large}. These models are shown to empower various stages of the design process, from ideation~\cite{jeon2021fashionq} and planning~\cite{zhong2024ai} to prototyping~\cite{kim2022stylette}, and have proven effective in augmenting creativity and improving efficiency~\cite{wu2021ai}.

The proven effectiveness of AI in supporting individual creators across the design lifecycle~\cite{frich2019mapping, wu2021ai, shin2024paper} motivates us to explore how these technologies can address broader design challenges. Current AI tools powerfully augment tasks like ideation and prototyping for the individual~\cite{gonzalez2024collaborative, kim2022stylette}, yet the vital feedback loop involving prospective stakeholders often remains unaddressed by these systems. Our work extends the application of AI in creativity-support tools by introducing virtual audience-driven persona agents, informed by the marketing document, that provide context-aware feedback and facilitate discussions, embedding the potential audience perspective directly within the individual designer's canvas.

\subsection{Generative Agents}

Recent work in HCI has explored how personas can be leveraged in the use of generative AI to offer more tailored support. For instance, Ha \etal{}~\cite{ha2024clochat} investigated how individuals can personalize AI personas to enhance human-AI interaction experiences. This concept has been adapted by practical applications, such as customer service~\cite{shi2024chops} and education~\cite{markel2023gpteach}, where personalized AI interactions have been shown to not only improve user experience but also increase the utility and acceptance of AI systems across various domains.

Recent research has expanded beyond individual human-AI interactions by exploring how generative AI can stimulate social interactions. Park \etal{}~\cite{park2022social} introduced the concept of populated prototypes to help researchers build social computing systems for evaluating social interactions, which has led to the development of generative agents---computational agents capable of simulating believable human behavior~\cite{park2023generative}. These agents can model diverse social scenarios with diversified persona agents, providing insights into human behavior and social dynamics. By leveraging these simulations, researchers have been able to create various scenarios to test hypotheses in controlled environments, minimizing constraints imposed by real-world interactions.

The use of generative AI to simulate social interactions holds promise beyond merely replicating human traits; it poses significant opportunities in simulating interactive behaviors in pragmatic tasks (\eg{}, software engineering~\cite{qian2024chatdev}). In our work, we extend these advancements by applying agents to the design collaboration process. By grounding agents' perspectives in real-world data widely used in the marketing sectors, with each persona agent providing unique perspectives, we address the challenges of poster design collaborations---facilitating ideation, fostering empathy-driven design, and ensuring inclusivity through a broad range of target audience perspectives~\cite{smeenk2019comparison}.
\section{Design Space}

To conceptualize our proposed idea of audience-driven persona agents for poster design feedback, we started by understanding and scoping their design space. Specifically, we explored the resources that would allow us to create personas based on the poster's target audience, reviewed the literature that could inform the design space for leveraging audience-driven agents for poster designing, and detailed the key takeaways by theme. Lastly, we built upon these initial explorations with our formative study to derive our design goals.

\subsection{How can we effectively establish target audience roles using real-world materials from the existing workflow?}\label{sec:brief}

To develop effective persona agents that enrich the poster design from the target audiences' perspectives, it is critical to establish each target audience's role. Prior work suggested that LLMs can be prompted to take on various backgrounds as a form of \textit{personas} to simulate their behaviors~\cite{park2023generative}. In this process, each persona agent needs a defined \textit{identity}, which encompasses a unique set of background, to create varying roles and perspectives on the target of the simulation---in this case, simulating target audiences of the marketing campaign. These elements can be formulated in the form of seed memory using persona description~\cite{park2022social, park2023generative}.

However, manually crafting personas and using them as seed inputs for agents is labor-intensive and often demands thorough user research~\cite{matthews2012designers}. One potential solution to address this is to use \textit{marketing briefs} (also known as design or creative briefs) as guidance. Marketing briefs are structured text documents that serve as the foundation for creative work~\cite{delash2022isosceles} and are widely used in the commercial marketing sector to develop playbooks that generate actionable insights~\cite{playbook}. In general, they provide a broad description of the audience that they are targeting, along with an overview of a marketing campaign's goals and marketing details (\eg{}, message to deliver, constraints)~\cite{ryd2004design}. A concrete example is illustrated in \autoref{fig:marketing-brief}, which outlines the details of an oral care brand's campaign---containing sections like the problem statements, target audiences, and goals (\ie{}, driving increased market share among young customers), which can serve as the foundation for the agents' persona. However, since these descriptions are providing a high-level overview of the target users and are not intended for crafting highly detailed profiles like personas, these audience summaries need to be refined and structured into personas that capture diverse audience segments and provide meaningful feedback.

\begin{framed}
\noindent
\edited{
\textbf{Takeaway 1}: Marketing briefs can provide foundational basis, such as overview of target audience and a goal, for establishing audience-driven agents. However, they need to be refined and structured to derive diverse and meaningful personas.
}
\end{framed}

\subsection{What are the usage patterns of design tools for collaborative feedback, and where could audience-driven persona agents help?}

Despite the lack of prior works directly exploring in situ feedback of the AI agents on the design canvas in poster design, prior works have explored the collaborative design feedback of goal-driven UX designing in corporate organization settings~\cite{feng2023understanding, o2018charrette}. We believed these works would provide a useful foundation for surfacing the user expectations of agents in a collaborative design tool, and defining the expected role of our audience-driven agents for design feedback.

In these works, the roles of stakeholders in collaborative design are characterized as \textit{author} and \textit{editor}, where the author takes the role of generating the initial draft of a design idea, which is then subject to revisions or comments from the editors. With this collaborative structure, editors are tasked with providing feedback on the shared canvas, often enabled by text-based comments for discussions and reasoning, visual annotations, and sticky notes. To support this process, various strategies have been proposed to facilitate effective collaborations~\cite{o2018charrette}, such as  \textit{tracking the changes and evolution of ideas}, \textit{taking notes on diverse opinions}, and \textit{summarizing the pain points}.

Throughout this feedback process, audience-driven persona agents may potentially be capable of providing diverse supports, yet it is essential to understand how and to what extent designers anticipate these design supports to be delivered within their modality to maintain their agency~\cite{chaudhry2024concerns}---which is yet unclear.

\begin{framed}
\noindent
\edited{
\textbf{Takeaway 2}: Collaborative design tools primarily facilitate feedback through author-editor roles, while it remains unexplored how the role and modality for audience-driven persona agents need to be formulated to provide in situ feedback without compromising user agency.
}
\end{framed}

\subsection{How is AI currently used to support poster design?}

Previous research in HCI and AI on poster design has primarily emphasized automation, framing the design process as the refinement of key components to create a poster (see \autoref{tab:design-recommendation}): (i) text (\eg{}, tagline, description), (ii) images, and (iii) themes (\eg{} layout, font style, embellishments)~\cite{guo2021vinci, lin2023autoposter, wang2024prompt2poster, chai2023two, li2023relation, vaddamanu2022harmonized}. In these works, the granularity of theme feedback varies; some recommend an entire theme based on the content~\cite{guo2021vinci}, while others focus on specific elements, such as structuring the layout~\cite{chai2023two, lin2023autoposter, li2023relation} or styling fonts for text components~\cite{vaddamanu2022harmonized} to ensure visual coherence within predefined sets of the other text, image, and theme elements.

\begin{framed}
\noindent
\edited{\textbf{Takeaway 3}: AI currently supports poster design primarily through automating the refinement of (i) text, (ii) images, and (iii) themes, with varying granularities of theme feedback---ranging from entire theme recommendations to specific element styling.}
\end{framed}

Motivated by these examples for designing an advertisement poster, in this work, we focus on the agents providing feedback of (i) text, (ii) image components, and (iii) theme as a unit of feedback.

\begin{table}[t]
    \caption{Examples of previously proposed approaches for leveraging AI in supporting poster design. These works mainly focused on automating the creation of missing poster components (\ie{}, texts, images, and themes) in alignment with the input poster components.}
    \footnotesize
    \begin{tabular}{lll}
    \toprule
    \textbf{Input(s)} & \textbf{Generated component(s)} & \\
    \toprule
    \leftcell{Text (tagline), image} & \leftcell{Theme (layout, background color,\\embellishment)} & \cite{guo2021vinci} \\
    \leftcell{Poster description,\\image, poster size} & \leftcell{Text (tagline), theme (layout)} & \cite{lin2023autoposter} \\
    \leftcell{Poster description} & \leftcell{Text (tagline), theme (layout), image} & \cite{wang2024prompt2poster} \\
    \leftcell{Image} & \leftcell{Theme (layout)} & \cite{chai2023two} \\
    \leftcell{Text (tagline)} & \leftcell{Theme (layout)} & \cite{li2023relation} \\
    \leftcell{Text (tagline), image} & \leftcell{Theme (layout, font style)} & \cite{vaddamanu2022harmonized} \\
    \bottomrule
    \end{tabular}
    \label{tab:design-recommendation}
    \Description{Table 1 summarizes previously proposed AI approaches for poster design. It has three columns: input(s), generated component(s), and citations. Each row describes a different method, for instance, showing that an input of text (tagline) and image can be used to generate a theme (layout, background color, embellishment).}
\end{table}

\subsection{Formative Interview Study}

Our literature review on the design space laid the groundwork for using persona agents as representations of target audiences. However, user preferences in constructing these agents, expectations of their behavior, and their practical application remain unclear. To probe this, we conducted a semi-structured, think-aloud interview study with designers. Participants were given materials that would serve as the foundation for building the agents intended to support design. They then engaged in a think-aloud process, providing insights on the end-to-end design procedure---from constructing these agents using the provided materials to the overall design process.

\subsubsection{Participants}

We recruited 8 professionals who had prior experience in advertisement poster design (FP1 -- FP8) from 3 design-focused communities. Our recruitment strategy sought diversity across multiple dimensions: years of work experience (ranging from entry-level to senior designers), domain area (\eg{}, enterprise marketing, retail advertising), as well as the professional role (\eg{}, UX researchers, marketing specialists, and communications managers). Our participants had an average of 4.4 years of experience ($SD=2.4$), and had experience creating posters for various advertising contexts---including industry events, marketing campaigns, and community events (see~\autoref{tab:participants}).

\subsubsection{Procedure \& analysis}

The interview sessions were conducted remotely on Microsoft Teams. We began each session by asking questions about their demographics and professional experience in poster design and relevant multimedia formats. Next, we presented each participant with a sample marketing brief, along with a simple poster draft containing a tagline, image, and descriptive text.

The study then proceeded through three key inquiry phases. (i) First, we asked participants to describe their process for identifying and constructing target audience profiles based on marketing briefs. (ii) Then, we explored how they would present feedback from the perspective of these diverse audience profiles. (iii) Finally, we investigated their strategies for reconciling potentially conflicting viewpoints when presented with multiple audience perspectives. Throughout the session, participants were encouraged to think aloud and explain their reasoning as they worked through these design challenges.

Each interview lasted approximately one hour, and participants received 25 USD as compensation upon completion. We then transcribed the recordings and conducted a thematic analysis~\cite{braun2006using}. Following a bottom-up approach, two authors independently identified themes from the transcripts before regularly meeting to discuss and refine them. Through four rounds of discussion, they reached a consensus on the final themes presented in Section~\ref{sec:study-results}.

\subsubsection{Study results}\label{sec:study-results}

Following, we detail the results from the thematic analysis of our interview sessions:

\begin{itemize}
    \item \textit{Steer attributes presented in the marketing brief to shape diversified audience groups and enable a broad understanding of target audiences.} All eight participants highlighted the importance of identifying key attributes emphasized in marketing briefs to generate audience representations that cover diverse cases: \textit{“We would break down the attributes in this brief to characterize the potential groups we should or could get feedback from.”}~(FP7) They noted that, by leveraging these attributes and creating variations using them, designers can ensure the resulting virtual audiences comprehensively represent diverse potential target audiences: \textit{“Knowing who I market to isn't always straightforward, I really dig deep into the product to characterize and steer them.”}~(FP5)
    \item \textit{Contextualize multi-level feedback within the design canvas.} Seven participants emphasized the dual importance of integrating feedback directly into the design space, while providing both high-level critiques and specific actionable suggestions. The combination of overarching insights and detailed recommendations is suggested as the method for providing both strategic direction and concrete implementation paths: \textit{“The type (FPrescriptiveness) of feedback I want depends (...) why I need it and who I get it from essentially says what form I want it in.”}~(FP5) Also, highlighting that poster elements exist across multiple layers, they emphasized the importance of visually mapping feedback onto corresponding components: \textit{“(because) there are many layers to designing a poster.”}~(FP3)
    \item \textit{Consolidate theme feedback instead of feedback for individual styles.} On top of these insights, all of the four participants who mentioned the delivery of theme feedback described the need for providing preferred theme-related feedback structured around broader themes rather than fragmented style suggestions (\eg{}, font, color scheme, layout). They noted that presenting cohesive examples of complete design themes helps designers recognize overarching patterns and make more holistic, intentional design adjustments that were verified by the previous designers, rather than reacting to scattered individual comments and having to reconcile them again:  \textit{“Before the nitty-gritty, you could say I'm vibe-designing (for the theme). I'm gonna check if it even looks right.”}~(FP8)
    \item \textit{Support reconciliation of conflicting viewpoints to generate insights that satisfy most audiences through discussions.} Acknowledging the conflicting perspectives that different agents might have, four participants emphasized the need for a structured approach to resolve these conflicts and align design changes with the majority of the target audiences' preferences, such as discussions or synthesis methods. This process is reported to potentially help designers create designs that accommodate diverse potential target audiences:
    \textit{“(If) no one (target user) has the perfect idea, we have to mix-and-match their feedback.”}~(FP6)
\end{itemize}

\subsection{Design Goals}
Following our exploration of prior works and our formative interview study, we set the design goals of a persona-driven collaborative assistant for poster design as follows:

\begin{itemize}
    \item[\textbf{DG1}] Generate diverse persona agents constructed from research on real-world audience data (\ie{}, marketing brief) by leveraging its key attributes to enhance audience diversity, enabling the simulation of real-time audience feedback within the design tool.
    \item[\textbf{DG2}] Enable agents to provide contextualized feedback on poster components (\ie{}, text, image, consolidated design theme) within the canvas, while offering both an inspirational insight and an actionable suggestion (\ie{}, actual modification to be applied) for feedback.
    \item[\textbf{DG3}] Support discussions among the persona agents and the user to reconcile conflicting design feedback through articulation and by reaching a conclusion that maximizes satisfaction among all agents concerned.
\end{itemize}
\section{The \system{} System}
Based on the design goals established from our exploration of the design space, we designed and developed \system{}, a canvas-based design system featuring audience-driven persona agents to assist designers in creating advertisement posters (see \autoref{fig:keyscreen}). In this section, we provide an overview of \system{}, a backend pipeline, and its frontend components.

\subsection{System Overview}

\begin{figure*}
    \centering
    \includegraphics[width=\linewidth]{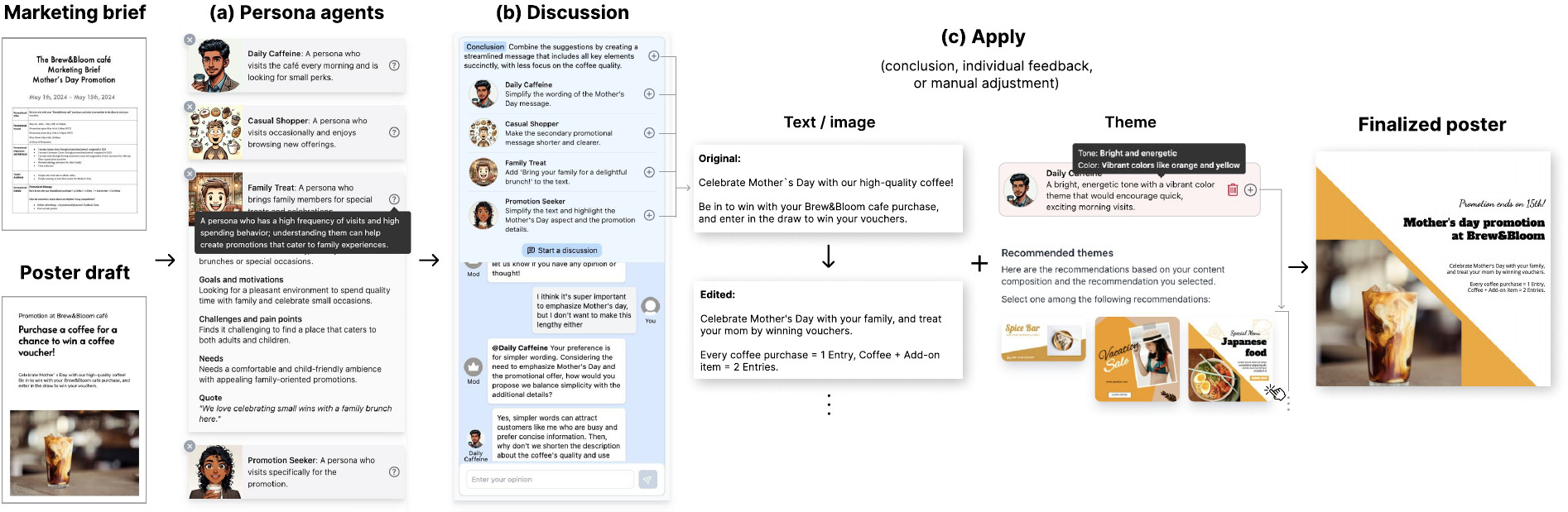}
    \caption{User interaction scenario using \system{}. With the marketing brief uploaded, the user is presented with (a) a list of personas that may suggest design feedback for their poster draft.  Then, the user can (b) discuss with these persona agents to reconcile viewpoints and reach a conclusion for (c) text, image, and theme, and apply them to the canvas to finalize the design.}
    \Description{Figure 2 is a flowchart illustrating a user interaction scenario in PosterMate. The process begins with (a) a marketing brief and a poster draft, from which the system generates four distinct AI personas. Next, (b) the user can initiate a discussion with the personas in a chat interface to reconcile conflicting feedback. Finally, (c) the user can apply the resulting conclusion or individual feedback to the poster's text, image, and theme, leading to a finalized poster.}
    \label{fig:order}
\end{figure*}

In the \system{} system, we define the user's role as an \textit{author} based on our design space exploration, while multiple persona agents play the role of collaborative \textit{editors} that provide feedback on edits and participate in a discussion to reach a conclusion.

The user first begins by drafting components (\eg{}, text, image) on the canvas screen, and uploads a marketing brief to generate persona agents. Based on both the persona descriptions and the user's initial design, \system{} then generates design feedback from each persona agent on the components and theme, both at a high level and with specific modifications. The user can choose to accept an individual agent's feedback, or initiate and engage in discussions with the persona agents to reconcile their diverse feedback for each design component.

\begin{figure*}
    \centering
    \includegraphics[width=\linewidth]{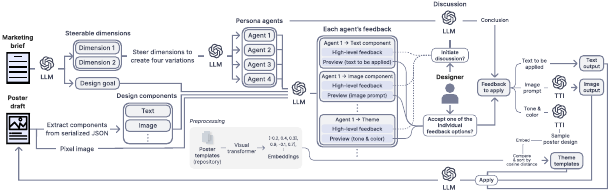}
    \caption{Pipeline of the \system{}-driven feedback and iteration process}
    \label{fig:pipeline}
    \Description{Figure 3 is a detailed pipeline diagram of PosterMate's feedback and iteration process. It shows how a marketing brief is used by an LLM to create four persona agents based on steerable dimensions. Concurrently, a poster draft is processed into its design components (text and image). Each agent then provides feedback on these components. The designer can either accept an individual's feedback or initiate a discussion moderated by a moderator agent to reach a conclusion. This outcome is then applied to the design, which is rendered back on the canvas for further iteration.}
\end{figure*}

Once the individual feedback or discussion outcome is accepted by the user, \system{} automatically integrates it into the poster design. If not, the user can manually adjust components, or initiate follow-up discussions to further refine the conclusion. Additional edits can be made as needed before finalizing the poster design. The overview of the user interaction scenario and the system pipeline is illustrated in Figure~\ref{fig:order} and \ref{fig:pipeline}, respectively.

\subsection{Data Preparation and Generative Models}

Recent advancements, particularly in LLMs extended into multimodal space, have shown a remarkable ability to generalize across tasks and handle multimodal inputs like text and images~\cite{achiam2023gpt}. This makes them well-suited for simulating design suggestions and facilitating discussions about visual posters based on verbal dialogue. To this end, \system{} employs a multimodal LLM to analyze poster designs and generate persona-driven agents. Leveraging two distinct representations of the poster---its pixel-based image and a serialized JSON structure---as input, \system{} aims to provide the model with a flexible and rich basis for the model to interpret and reason about the poster's visual appearance and underlying content layout effectively.

\edited{Similarly, \system{} leverages multimodal LLM to extract key information (\eg{}, goals, a broad description of the target audience; see \autoref{fig:marketing-brief}) needed for constructing agents from a marketing brief. Although \system{} assumes the use of marketing brief, it can also draw from other materials provided they are in text or image format and contain the relevant information.}

For generative tasks, \system{} uses OpenAI's GPT-4o~\cite{gpt4o} as the LLM and DALL-E 3~\cite{dalle3} for text-to-image generation. Our modular backend allows these models to be easily replaced with alternatives that can offer similar capabilities.

\subsection{Constructing Persona Agents from Marketing Brief (DG1)}\label{sec:construct_persona}

Following our formative study results, we decided to elicit attributes from the marketing brief, and leverage these as \textit{dimensions} for constructing persona agents that simulate a set of target audiences. Prior literature in business and organization studies suggests that a 2×2 matrix with the combination of varying levels within each of the two dimensions is often used as a structured yet flexible framework for quickly mapping the problem and design space~\cite{pollock2012give, lowy2010power, 2x2edition}. Following this, we decided to formulate four persona agents with diverse representations using the combinations of two elicited dimensions. Through positioning personas along these dimension pairs,  we believed that the system can enhance understanding of target audiences and design opportunities across various potential marketing targets. In addition, choosing two dimensions would prevent users from being overwhelmed by an excessive number of agents and benefit in terms of computational scalability---adding an additional dimension will multiply the number of personas by a factor of 2, which could affect the efficiency of downstream tasks like discussions and component feedback significantly.

To elicit dimensions from the marketing brief, an LLM is first prompted to return two dimensions that could be varied given the marketing brief, without contradicting audience characteristics defined in the marketing brief. Specifically, we instructed the model to identify the dimension (i) that is steerable (\ie{}, can be adjusted to higher or lower values) and (ii) whose extremes would remain compatible with any preexisting target user descriptions in the marketing brief to avoid conflicts with the original marketing goal. If such steerable dimensions are identified from the brief, the model is instructed to prioritize those; otherwise, it generates two steerable dimensions contextually relevant to the brief's target campaign, whose combinations of their extremes do not contradict the brief's contents. For example, if the brief targets the customers of a shopping mall, the generated dimensions are returned as \textit{frequency of visits} and \textit{engagement level}. Then, the model generates four personas based on the combination of two extremes of these axes (\eg{}, frequent shopper vs. occasional shopper × passive browser vs. active shopper) with specifications from the marketing brief. An example of this process is illustrated in \autoref{fig:marketing-brief}.

Then, with these four variations, the model constructs each persona formatted in a fixed qualitative description with the following elements (see \autoref{fig:order}a): (1) \textit{name} (a two-word description of the persona; \eg{}, \texttt{Trend Skeptic} for the persona with low fashion sensitivity), (2) one-line \textit{summary} of the persona description, (3) \textit{background} that exemplifies the persona traits, (4) \textit{goal/motivation}, (5) \textit{challenge/pain point}, (6) \textit{need}, (7) one-line \textit{quote}, and (8) one-line \textit{rationale} (how the persona's perspectives could contribute to the poster design). Users can also provide manual details as inputs to add personas apart from the briefs to ensure user agency. Besides, we generate an avatar based on the name of the persona to increase users' social affordances for co-presence~\cite{feng2023understanding}. Lastly, we instructed the model to elicit a high-level goal of marketing from the brief, to later guide the persona agents' feedback not to deviate from the primary objective of the marketing.

\subsection{Generating Design Component Feedback from Persona Agents (DG2)}

After constructing the persona descriptions, \system{} leverages an LLM to have each persona agent provide feedback to the components (\ie{}, text, image, theme) of the poster (see \autoref{fig:order}c). In our design space exploration, we identified the importance of multi-level feedback---providing both high-level feedback and actionable suggestions. To support this, the system provides both high-level feedback (\ie{}, the persona agent's opinion) and preview (\ie{}, an illustration of how the component will be actually modified once accepted).

The model is first prompted with a serialized JSON format exported from the canvas tool (\ie{}, Polotno REST API), as well as its rendered format as a pixel image, to assist the model in understanding the exact contents and its representation in the poster. Given this input, the goal of the poster design, and the persona details, the model provides feedback from each persona agent with the following structure for each data type, respectively:

\subsubsection{Text feedback}
\begin{figure}[t!]
    \centering
    \includegraphics[width=\linewidth]{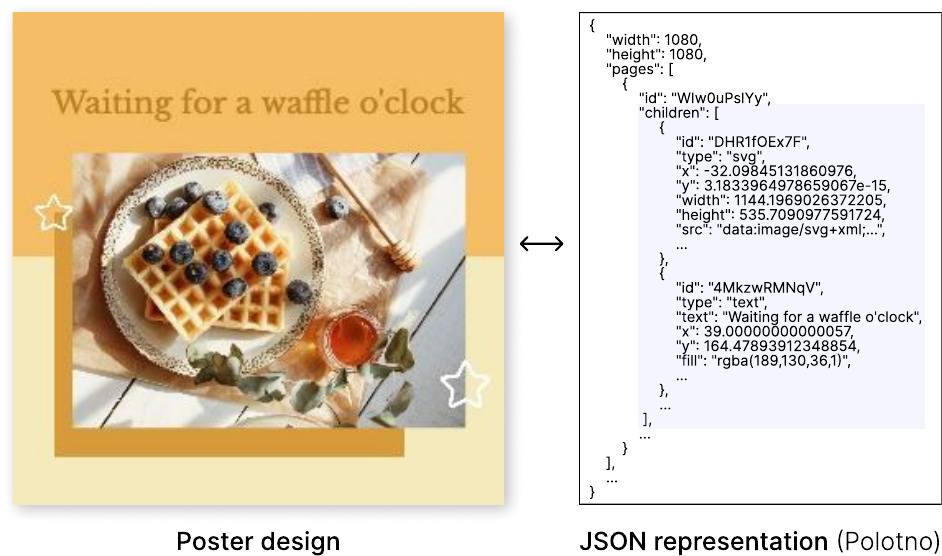}
    \caption{JSON representation of the poster design in Polotno~\cite{polotno}, where every element is saved in a \texttt{children} entity of the JSON (blue box). The canvas supports the import/export of the poster through JSON, enabling programmatic manipulation.}
    \Description{Figure 4 displays the data representation of a poster design within the system. On the left, an image shows a poster template. On the right, a code snippet shows the corresponding JSON representation, where each visual element on the poster, such as text and images, is defined as an object with properties for type, content, position, and size. A blue box highlights the children array in the JSON, which contains these element objects.}
    \label{fig:json_representation}
\end{figure}

For text components, each feedback from the persona agent is returned as a JSON object, which includes: (i) an identifier for the target text component, as defined in the Polotno-exported JSONs (see \autoref{fig:json_representation}). Also, in line with our design goal of supporting multi-level feedback, the JSON includes: (ii) the persona agent's opinion on how the text could be changed to better align with their expectation (\ie{}, high-level feedback), and (iii) a content of the text after which that feedback is applied (\ie{}, preview)---the structure of which allows for both offering high-level insight and actual modification to be made.

\subsubsection{Image feedback} 
For image components, an LLM returns for each feedback a JSON consisting of (i) an identifier for the target image component, (ii) the persona agent's opinion on how the image could be changed to better align with their expectation (\ie{}, high-level feedback), and (iii) a one-line text description illustrating the image to be applied once the feedback is accepted (\ie{}, preview). The content of (iii) is later used by \system{} as a prompt for text-to-image generation once selected. Although this description can be used to either suggest an image from the database by matching it with the closest image captions (\eg{},~\cite{zhou2024epigraphics}) or generate a new image using a text-to-image model, we focused on text-to-image generation for its scalability, motivated by the prior work~\cite{shin2024paper}. Regardless, this approach could easily be substituted with alternative techniques, such as an image retrieval task.

\subsubsection{Theme feedback}\label{sec:theme-feedback}

A poster theme is a visual-textual presentation that contains varying combinations of tone and color, along with the associated font styles and embellishments~\cite{guo2021vinci, lin2023autoposter, li2020attribute, o2014learning}. Reflecting the close interconnection between these thematic elements~\cite{guo2021vinci}, participants in our formative study expressed a preference for receiving consolidated feedback on the overall theme rather than separate feedback for each individual style element. Following this, we decided to leverage a template database to enable more centralized and intentional reasoning, instead of providing feedback for piecemeal adjustments. More specifically, our system utilizes Polotno's template library\footnote{https://polotno.com/docs/templates-library}, which contains over 1,000 public examples, to recommend complete poster themes. We began by preprocessing these themes by embedding each using a vision transformer model (\ie{}, ViT-base-patch16-224-in21k~\cite{dosovitskiy2020image}), and storing them in a vector storage.

Then, when the \system{}'s agent provides a theme feedback, an LLM is prompted to return a JSON consisting of (i) the persona agent's one-line opinion of the mood that they want the theme to deliver (\ie{}, high-level feedback). To facilitate querying and retrieval of the theme, this mood is represented by (ii) two theme descriptors---\textit{tone} and \textit{color} (\ie{}, preview), which will be used as a preview of the theme once accepted. Once the user accepts the feedback, the system employs a text-to-image model to create a poster design that reflects the corresponding tone and color, which is then converted into an image embedding using the same embedding model as our preprocessing. Then, \system{} calculates the cosine similarity between the generated design image and each of the templates, and sorts the templates by similarity order. This results in a list of theme templates sorted by relevance based on the feedback.

\subsubsection{Application of the feedback to the canvas}

For text and image feedback, once the user accepts particular feedback, with the associated component identifier, the system first locates the target component from the JSON representation of the previous poster design (see~\autoref{fig:json_representation}). Then, for text, the system updates the corresponding component's \texttt{text} property (which specifies the text to be displayed) by replacing the value with the preview content of the agent's feedback. Similarly, for images, it replaces the \texttt{src} property, which holds the image source, with the URL of the image generated from the image feedback.

For the theme, once the user selects a particular theme from the list, the system takes the following steps to apply it to the previous poster design. First, given JSON representations of both the previous poster design and the selected theme template, the model is prompted to map the original components (\ie{}, texts, images) to the selected template with a corresponding text/image element. If the template contains extra components or is missing some, it adjusts by adding or removing them accordingly. In this process, During this process, the system temporarily stores theme embellishments (\ie{}, SVGs) in a separate array, along with their z-index values. This ensures the visual hierarchy is preserved when reintegrating these elements into the updated JSON, especially since such embellishments are often represented as raw data, which may exceed the context window of the LLM.

Then, motivated by prior work demonstrating LLMs' potential to understand and manipulate the numeric positions of graphical elements~\cite{masson2024directgpt}, we designed \system{} to take both a pixel-based image and a JSON representation of the poster as input, enabling it to detect and resolve visual overlaps. Once the inputs are provided, the model checks if (i) repositioning (\ie{}, adjusting x/y coordinates) or (ii) resizing (\ie{}, adjusting width, height, or text size) of any components is needed to prevent visual overlap. If adjustments are required, it also returns the IDs of the affected components along with a one-line description of the necessary change, based on its interpretation of the provided pixel image. Then, based on the description, the model updates the relevant component's position or size in the JSON accordingly. When no further changes are required, the poster design is rendered on the canvas, allowing the user to correct any remaining issues from the model's iteration and complete the design.

\subsubsection{Guardrails for ensuring alignment between the original marketing brief and the feedback}
Since the marketing brief serves as the original guideline, it is essential to ensure that the feedback generated by the persona agents aligns with the original brief. To achieve this, we implemented several guardrails for each feedback generation to ensure that feedback aligns closely with the persona agent's specifics and the brief.

First, we directed the model to align with the primary objective of the poster design, as outlined in Section~\ref{sec:construct_persona}, to ensure that the persona agents' feedback does not deviate from this goal. Additionally, each instance of feedback generation by the LLM was informed by the raw marketing brief data, enabling the persona agents to incorporate specific details from the brief into their feedback. Lastly, inspired by prior works demonstrating how returning rationale can improve accuracy~\cite{li2024can, tsai2024leveraging}, we prompted the system to generate both the rationale (\ie{}, ground) as well as the output, based on the provided persona details, the goal, and the marketing brief details.

\subsection{Discussion on Component Feedback (DG3)}
Our formative study revealed the need for supporting reconciliation when the agents have conflicts in feedback on the design component. To support this, \system{}, upon user request, prompts an LLM to perform a series of moderation steps designed to resolve design feedback conflicts through a panel discussion-style approach (see \autoref{fig:order}b). Specifically, a moderator agent takes the role of asking questions and controlling the conversation flow among the persona agents until a conclusion is reached as follows:

\subsubsection{Grouping feedback by target component \& detecting conflicts}
To identify conflicting pieces of feedback for the same component, the system first groups feedback by the component identifier. Then, an LLM is instructed to detect a set of conflicting feedback that has distinct focuses within each group and generate a one-line summary of the conflict.

\subsubsection{Generating and answering thought-provoking questions}\label{sec:answer_question}

Previous research has highlighted the effectiveness of using thought-provoking questions to enhance the process of critiquing and iterating on design feedback~\cite{crain2017share}. Building on this insight, we designed the system to provide users with an option to run a panel discussion for each conflict, where the moderator asks each persona agent to respond to an open-ended, thought-provoking question.

First, the moderator asks the user if they have any opinion (\ie{}, user comment) about the conflict to center the discussion on. Then, based on the conflict summary, persona details, and the user comment (if provided), the moderator generates a thought-provoking question for each persona agent, requesting them to articulate their motivation and potential ways to narrow down the conflict with the user comment serving as a constraint, to facilitate a conclusion-building discussion. Similar to what we did for generating feedback, in this process, the model is informed with the marketing brief data along with the goal of poster design to ensure that the discussion does not deviate from the marketing objective and details.

Each persona agent is then prompted to respond to their specific questions in an open-ended manner, with instructions to remain open-minded and collaborate towards finding a compromise to resolve the conflict. Since the responses only focus on the questions, we opted to generate the answers among the persona agents in parallel, instead of alternative approaches that could be used to perform similar tasks (\eg{}, chain-of-thought prompting~\cite{wei2022chain}), to reduce latency.

\subsubsection{Resolving potential conflicts \& drawing a conclusion}

Once the response from each persona agent is collected, we prompt the moderator agent to draw a conclusion with the same data structure as the individual feedback -- (i) a component identifier, (ii) a summary of the conclusion, and (iii) a preview for the conclusion. Here, we defined conclusion as the final design feedback, formulated by the moderator, that accommodates individual responses from each persona agent while satisfying the user's comment (if provided).

More specifically, to address potential conflicts, \system{} follows a two-step process. First, at the individual feedback level, each agent is encouraged to seek areas of compromise and answer the moderator's question, as detailed in Section~\ref{sec:answer_question}, to maximize the likelihood of conflict resolution and reach an outcome that satisfies most agents later on. Second, if the agents' differences remain unresolved even after their responses, the moderator may selectively omit certain portion of some individual agents' perspectives to achieve a conclusion that meets the perspectives of the majority.

Then, the user can either accept it and apply the feedback directly to the canvas, or provide a subsequent comment to the conclusion so that the moderator will iterate on the discussion by generating thought-provoking questions for the persona agents again. An example of conflict resolution through discussion is illustrated in Appendix~\ref{sec:example_discussion}.
\subsection{Frontend Interface}\label{sec:frontend}

The frontend user interface of \system{} is built as a web application on SvelteKit~\cite{sveltekit}, a JavaScript-based web app framework. \system{}'s frontend is designed to communicate with Polotno API~\cite{polotno} to provide a canvas space for creating and editing the poster content and interface with our discussion feature. Every generative AI computation (\ie{}, LLM/VLMs, text-to-image models) is performed in our Python backend server, where the API calls for feedback and discussions are sent to our server via FastAPI. The user can begin interacting with the interface by uploading their marketing brief.

When designing the \system{} interface, we made several design choices to ensure that designers could maintain their agency and engage actively throughout the process of enhancing the poster design. First, at the beginning and end of each discussion, \system{} actively encourages designers to iterate with agents by asking them to provide opinions in the free-form text, if any. They are also encouraged to make adjustments directly on the canvas screen if required, allowing flexibility beyond the system's design feedback. Additionally, we sought to help designers guide their design choices by providing interpretive layers (\eg{}, previews, rationales provided through dialogue), supporting their agency of decision-making rather than relying entirely on automation.

Each view that consists of \system{} interface, as well as its corresponding design goal, is as follows:

\subsubsection{List of persona agents view \textnormal{\textbf{(DG1)}}}
Once the user uploads a marketing brief, the system generates multiple persona agents that can be viewed on the side panel (see \autoref{fig:keyscreen}-\circledigit{1}). Each persona agent view contains a cartoonistic avatar generated by a text-to-image model, its name, and the summary of its description. Users can also hover over the persona agent to view the one-line rationale on how the persona's perspectives could contribute to the poster design, and click to expand and view the details of the persona agent, as detailed in Section~\ref{sec:construct_persona}.

\subsubsection{Design feedback units \textnormal{\textbf{(DG2)}}}
After the list of persona agents is finalized and the user proceeds, the design feedback for each component from the persona agents is displayed on the side panel, where the set of feedback from multiple persona agents on the same design component is organized as a \textit{design feedback unit} (see \autoref{fig:keyscreen}-\circledigit{2}). To achieve our goal of visually mapping the design feedback onto its corresponding component within the canvas, the corresponding component of each unit in the canvas is highlighted once the user hovers over the unit for feedback on text/image components.

Within each unit, a concise summary of the conclusion or conflict among feedback that consists of the unit is displayed at the top. Users can view each individual persona agent's feedback below the summary, and the system also presents a tooltip for each feedback and conclusion that presents a preview of the feedback, which appears on hover. For text feedback, the preview is the actual text to be applied, and for image feedback, the preview represents a one-line illustration of the expected output. For theme feedback, the preview is a text description of the tone/color of the theme if it is applied, which will in turn result in a sorted list of theme templates when selected. Also, each unit's background color represents the current state (\ie{}, blue---conclusion reached or user applied one of the individual feedback vs. red---conflict unresolved).

\subsubsection{Chat interface for discussion \textnormal{\textbf{(DG3)}}}
Each unit also contains a chat interface, which is designed to run a discussion. The chat interface is expanded when the user clicks ‘Start a discussion’ button to initiate the discussion (see \autoref{fig:keyscreen}-\circledigit{3}). Once the discussion starts, the conversation between the moderator and the persona agents is displayed in the chat interface. The user can also share any comment before the discussion begins to center the discussion on. Then, once the conclusion is reached, the unit's conclusion displayed on the top will be updated accordingly. If the user wants to run a discussion with additional feedback, they can provide the comment after the conclusion is derived, prompting another round of discussion.
\section{User Study}\label{sec:user_study}
To explore the use of \system{} and understand how its approach of involving audience-driven persona agents in design feedback impacts designers, we conducted a user study with 12 participants.

\subsection{Participants}

We recruited participants from two design-focused communities and used snowball sampling to reach additional individuals. To ensure relevant feedback, we required participants to have prior experience in poster design, with at least one poster created previously. As a result, 12 participants were recruited; 8 self-identified as female and 4 as male, with an average age of 26.3 years ($SD = 4.4$).

\begin{figure*}
    \centering
    \includegraphics[width=.75\linewidth]{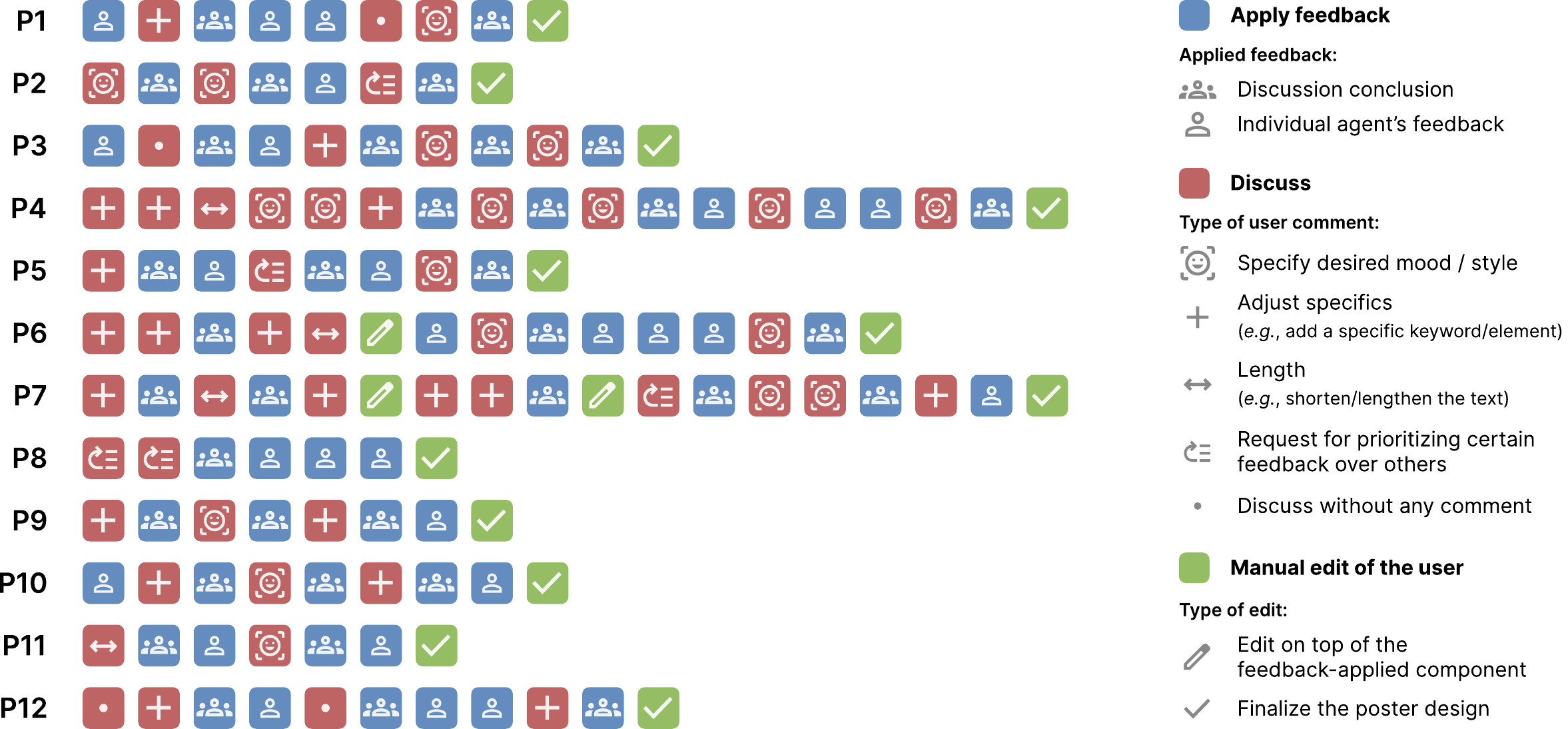}
    \caption{Sequence of the participants' interaction with \system{} in our user study. They actively collaborated with the persona agents to enhance and finalize their design of the advertisement poster.}
    \Description{Figure 5 is a chart that visualizes the sequence of interactions for 12 study participants (P1 through P12) using PosterMate. Each row represents a participant's workflow over time. A legend explains the symbols for different actions: applying feedback (from a discussion conclusion or an individual agent), starting a discussion (with different types of user comments), making manual edits, and finalizing the poster. The chart demonstrates that participants actively collaborated with the agents through a mix of discussions, direct feedback application, and manual adjustments to finalize their poster designs.}
    \label{fig:interaction_sequence}
\end{figure*}

Of all, 4 participants reported that they had previously designed posters less than 5 times, 6 had designed 5-10 times, and 2 more than 10 times. Participants had varying levels of design experience; 3 self-identified as novice, 5 as intermediate, and 4 as expert or as having/currently pursuing a professional design degree.

\subsection{Methodology \& Procedure}

Our study was conducted remotely via Zoom. After obtaining consent from each participant to record the interview, we provided a brief overview of our system and guided them through a tour. Then, each participant was provided a link where they could interact with our system interface.

Each participant was randomly assigned one of two marketing briefs (one for a café promotion marketing and the other for a sports brand advertisement; see \autoref{fig:brief} in Appendix~\ref{sec:study_details}), and they were first asked to familiarize themselves with it. Participants then uploaded the marketing brief to \system{}. During the study, they were asked to generate persona agents without customizing them to reduce the potential noise from customizing them and enhance the understanding of our automatically created personas. Then, participants were familiarized with the generated persona agents.

\edited{Following this, participants were tasked with iteratively improving a poster draft by freely interacting with \system{} to finalize their themes, reflecting the agents' intended role of supporting refinement based on our design space exploration. To facilitate this, the research team created two initial poster drafts, each featuring basic texts and an illustrative image (see \autoref{fig:draft} in Appendix~\ref{sec:study_details}), drawing from the two marketing briefs outlined in Section~\ref{sec:user_study}.} In this process, we employed the think-aloud method, where each participant was asked to verbalize their thoughts throughout the study, to capture their thought processes while interacting with \system{}.

After finalizing their posters, participants were provided with a survey link to answer questions using a 5-point Likert scale. The survey included (i) technology acceptance based on a subset of the Technology Acceptance Model~\cite{venkatesh2008technology}, which assessed perceived ease of use, perceived usefulness, and intention to use, as well as output quality, to evaluate the potential future adoption of \system{}. Additionally, since \system{} transcends information across modular components (\ie{}, marketing brief → persona agents → feedback on components → discussion (with user comments, if any) → conclusion), we also measured (ii) perceived logic~\cite{hwang2021ideabot} to determine how well the outputs from each step aligned with the preceding information. We then conducted an interview in which each participant was asked to describe their overall perception of our system, how they envision using our system in their future designs, and any future enhancements.

\sloppy
Each study took approximately an hour. Once the study was complete, each participant was compensated with a 25 USD gift card for their participation, and we transcribed the interview responses. We employed a thematic analysis~\cite{braun2006using} to analyze interview responses, where two authors initially reviewed the responses and identified themes independently. They then held regular meetings to discuss the emerging themes, finalizing them after four sessions. In Section~\ref{sec:user_study_results}, we refer to each participant as P1~--~P12.

\subsection{Results}\label{sec:user_study_results}

\subsubsection{User perception towards \system{} \& usage pattern}

\begin{figure*}
    \centering
    \includegraphics[width=.8\linewidth]{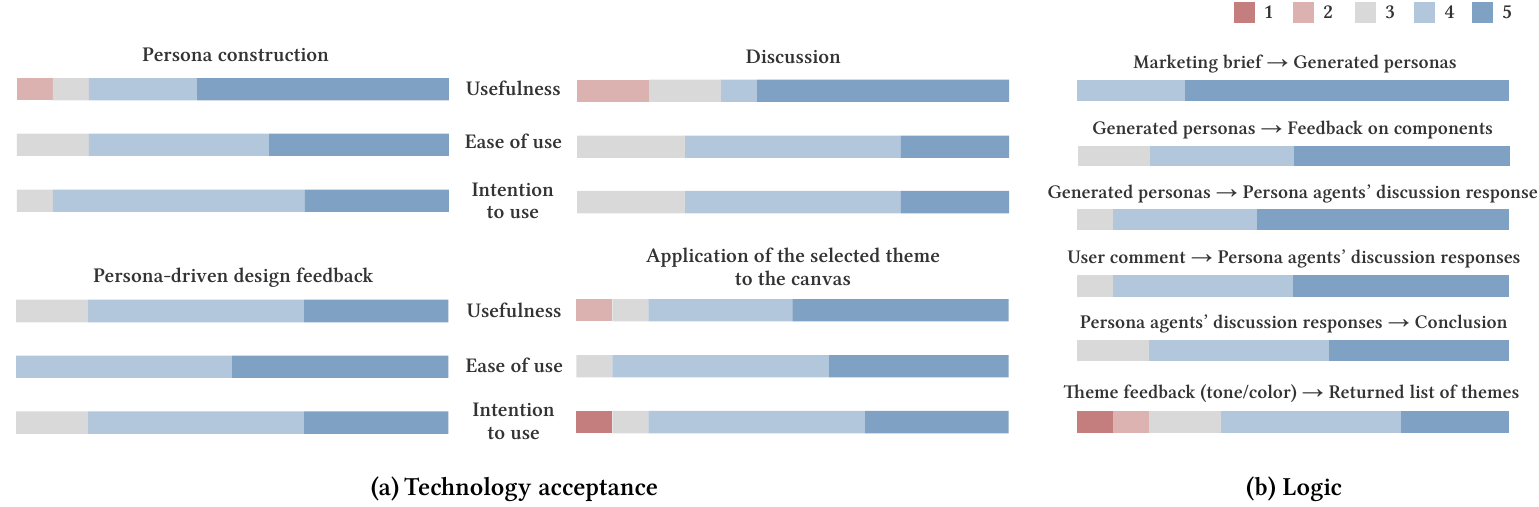}
    \caption{Survey results from our user study measuring (a) technology acceptance and (b) logic}
    \label{fig:quant_user_study}
    \Description{Figure 6 presents survey results in two parts. Figure 6a shows four stacked bar charts on technology acceptance for different system features. The charts measure usefulness, ease of use, and intention to use on a 5-point Likert scale, with results being positive. Figure 6b shows six bar charts on the perceived logic of the system's workflow, rating the connection between steps like "marketing brief to generated personas". These results are also positive, indicating users found the system's process logical.}
\end{figure*}

During our user study, participants actively used the discussion feature of \system{} to get feedback on the poster components. As illustrated in~\autoref{fig:interaction_sequence}, each participant ran 4.3 discussions ($SD=2.5$) on average to iterate on the design components and the theme, in order to modify the canvas contents and theme iteratively to revise the poster design. Our survey revealed that participants generally perceived each feature of \system{} as both useful and easy to use, and were willing to utilize the system in their future designs (see \autoref{fig:quant_user_study}a)\edited{, suggesting the potential efficacy of \system{} for real-world poster design scenarios}. Additionally, the participants reported that the generated personas aligned well with the marketing brief (see \autoref{fig:quant_user_study}b), demonstrating the reliability of constructing persona agents based on the briefs.

\begin{figure*}
    \centering
    \includegraphics[width=.7\textwidth]{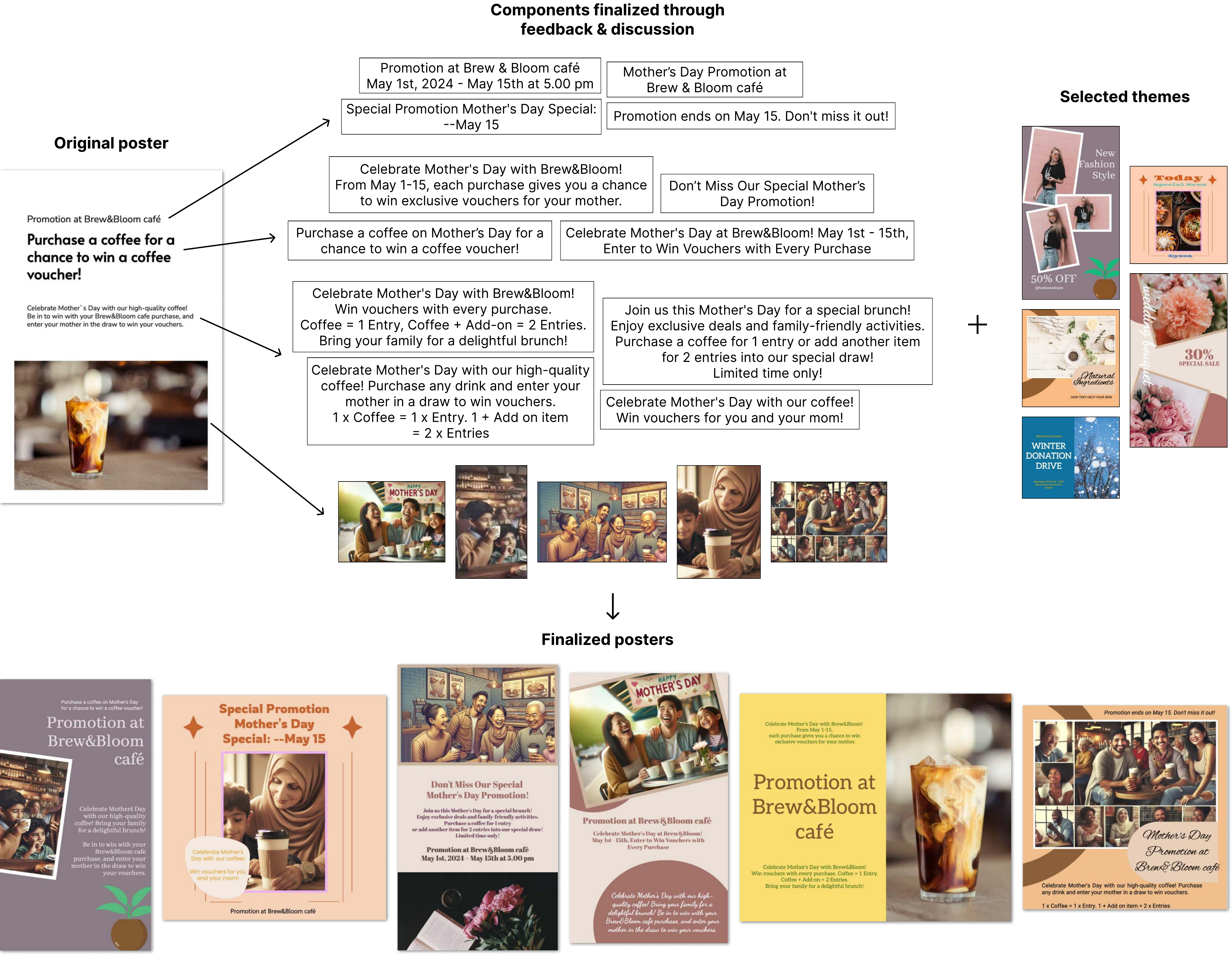}
    \caption{End-to-end illustration of how the participants assigned with the first marketing brief created posters}
    \label{fig:poster_composition_1}
    \Description{Figure 7 is a diagram illustrating the end-to-end poster creation process for the first marketing brief (a café). It starts with an original poster template. It then shows a table of components finalized through feedback and discussion, which are different text suggestions, and a grid of selected themes, which are visual style templates. The bottom of the figure displays a gallery of unique, finalized posters created by study participants, demonstrating the variety of outcomes produced using the system.}
\end{figure*}

\begin{figure*}
    \centering
    \includegraphics[width=.7\textwidth]{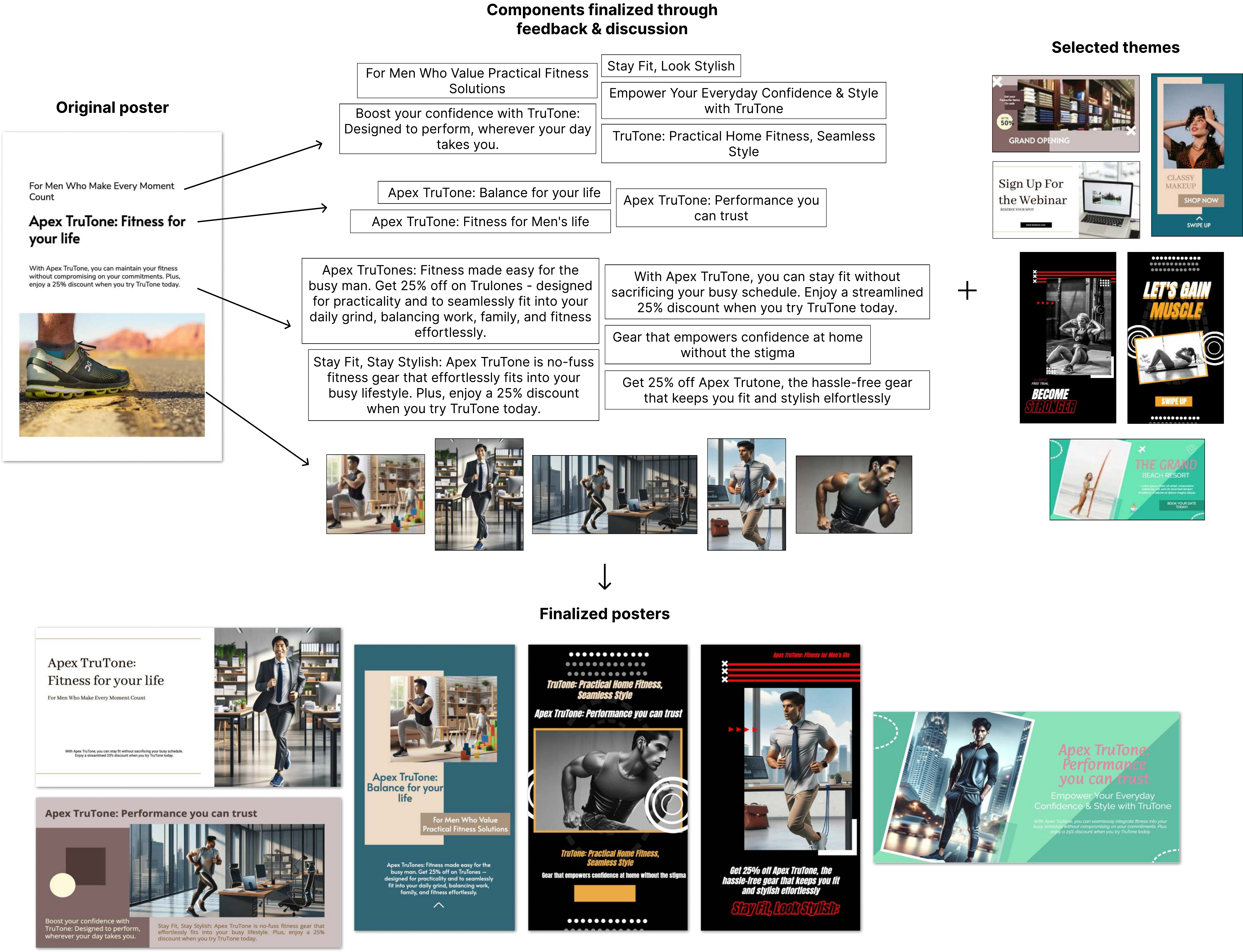}
    \caption{End-to-end illustration of how the participants assigned with the second marketing brief created posters}
    \label{fig:poster_composition_2}
    \Description{Figure 7 is a diagram illustrating the end-to-end poster creation process for the first marketing brief (a sports brand). It starts with an original poster template. It then shows a table of components finalized through feedback and discussion, which are different text suggestions, and a grid of selected themes, which are visual style templates. The bottom of the figure displays a gallery of unique, finalized posters created by study participants, demonstrating the variety of outcomes produced using the system.
}
\end{figure*}

\autoref{fig:poster_composition_1} and \ref{fig:poster_composition_2} illustrate the process by which participants interact with \system{} to finalize the posters, along with their intermediate steps. One notable observation was that, \system{} helped most participants to revise the contents of the poster components to highlight key information/message originally emphasized in the marketing briefs yet overlooked on the original poster (\eg{}, promotional details for the first brief, flexible training schedules from the second brief), where we could identify a similar pattern from the image (\eg{}, an illustration of a mother for the first brief, an illustration of a man training in daily surroundings for the second brief).

\edited{Additionally, as detailed in Section~\ref{sec:frontend}, we positioned \system{} as a supportive design assistant rather than an automated design tool. Reflecting this intent, we observed that participants proactively engaged with the agents' feedback through iterating, editing, or disregarding it at their own discretion, as illustrated in~\autoref{fig:interaction_sequence}. For example, P6 appreciated the tagline from the discussion conclusion, yet later found a particular word unnecessary; on top of the accepted conclusion, they made their own edit to finalize the text. Similarly, P7 initiated a discussion about the mood conveyed by an image; upon reviewing the discussion, they thought aloud that the issue might be tied to a specific part of the image and decided to apply the mood on that part only, prompting a new round of discussion. These examples show how \system{} supports the poster design process while the designer remains actively in control of the iterations.}

\subsubsection{Designer's interaction using \system{} to finalize the poster design}
\system{} enables users to engage directly with persona agents by actively participating in discussions. In line with our design goals, participants in our user study demonstrated several interaction patterns with \system{} to refine the design and finalize the poster.

For example, with the main conclusion in mind, some participants asked the agents to adjust the text length to their preference while preserving its message (\eg{}, \textit{“It is great, but can you make a shorter version?”}). They also requested adjustments of the specific phrase/element in the text/image (\eg{}, \textit{“I like the suggestion but is there any way to blend that with ‘men’ keyword?”}), emphasizing a certain mood or style (\eg{}, \textit{“Can you make it more modern, like black and white?”}), and prioritization of certain feedback over others (\eg{}, \textit{“Prioritize \texttt{agent\_name\_1} and \texttt{agent\_name\_2}'s opinions”}). In a few cases where the edits they wanted to make were minor (\eg{}, adding capitalization), we observed that they manually applied these edits after applying the feedback or conclusion from the discussion to the canvas. The full sequence of these participant-driven interactions leading to the final poster design is shown in~\autoref{fig:interaction_sequence}.

From our qualitative analysis, we identified the factors that contributed to \system{}'s effectiveness, as well as participants' perspectives on its future use and potential enhancements:

\subsubsection{Recognizing \& considering overlooked perspectives through receiving feedback from diverse persona agents} 

Instead of focusing on a single unified target audience, \system{} allows users to gain insights from a variety of potential poster readers as a form of persona agents. \edited{Participants noted the value of \system{}'s multi-agent feedback, highlighting that it helped them consider perspectives they might have otherwise overlooked in creating posters by comparing with their prior design experiences.} When designing in isolation, participants mentioned that they often found themselves overly focused on a specific use case or target audience, which limited their creativity and led to a narrow design scope. In contrast, through using \system{}, participants mentioned that the multiple persona agents that represented a spectrum of users allowed them to step outside of their initial assumptions and consider a wider range of potential needs and scenarios. Consequently, the design in turn could better capture overlooked points from the marketing brief and become more inclusive for broader audiences: \textit{“If I were to do this alone, I think I would just have been very focused on just one idea. But, you know, having other ideas like opinions kind of helped me like, oh, maybe I should tweak it into that way kind of thing.”}~(P2)

\subsubsection{Efficient prototyping through contextualizing feedback with previews}

Participants mentioned that the preview of the feedback made the decision-making process highly effective, without having to actually wait for the system to apply changes to the canvas. By simply hovering over the options to visualize the potential change to be made, participants reported that they could make faster and more informed decisions without experiencing excessive cognitive load: \textit{“I can see directly how they will be applied. That's like, I can see a semi-final version of the prototype already.”}~(P4)

Aligning with these results, participants commented that images and themes could be contextualized with the same modality, by offering the actual image/theme as a preview promptly. Currently, our system supports previewing feedback through textual descriptions only for images and themes due to computational efficiency. Participants suggested that the system could provide a more powerful visual stimulus that would assist in decision-making more effectively: \textit{“The preview (for text feedback) helped me because I’m a visual learner (...) if there's like a feature of an example image from (image feedback of) each persona, it'd be better.”}~(P3)


\subsubsection{Discussion helped understand motivations and inspirations behind persona agents' feedback} 

\sloppy
To finalize a component with diverse viewpoints, it may be helpful, yet challenging, to understand the motivations behind each feedback. In this context, \system{}'s moderator prompts each persona agent to respond to thought-provoking questions. Participants noted that this helped them to clarify the motivations behind their suggestions and significantly enhanced their decision-making for finalizing the decisions\edited{, suggesting how discussions supported by \system{} influenced poster creation by making decision-making more explainable}: \textit{“Often times systems like these lacked the amount of detail to inform me why it's useful. So although I would take that feedback, oftentimes I had to do much more to make sure that the recommendation was correct (...) That said, I think the fact that this (\system{} system) provided the feedback with different perspectives is useful.”}~(P8)

Meanwhile, some participants observed that such \textit{sensemaking} process could be made significantly more efficient with enhanced visualizations. Currently, the system displays the persona agent's responses as plain text dialogue. However, some participants pointed out that the text occasionally becomes lengthy, and incorporating visual elements, such as bullet points or the summary of the discussion, could facilitate the process: \textit{“I have to read it (discussion) over by scrolling. So it may be better if it highlights or gives me the summary of their opinions on how they got to the conclusion.”}~(P9)

Additionally, we acknowledged the necessity of equipping participants with a method to leverage intermediate ideas derived from persona agents' responses during discussions, since the discussions not only explained the design but also inspired users to produce more ideas. For instance, P7 noted that, after reviewing the motivation and an alternative idea presented by a persona agent, they unexpectedly gained valuable insight from the persona agent's dialogue. However, the current system does not support the real-time implementation of this intermediate feedback from an individual persona agent, requiring participants to manually apply the idea based on the persona agent's dialogue. This highlights the need for a mechanism that enables participants to effectively integrate intermediate insights within the conversational context.

\subsubsection{Integrating \system{} in the real-world workflow} 

After using \system{}, participants envisioned several use cases where the system would be well-suited. On the one hand, participants highlighted the potential effectiveness of \system{} in supporting fast-paced environments where dedicated design teams for user research are unavailable, or where novice designers struggle to understand the needs of target users. In these constrained situations, they viewed the idea of persona-driven collaborative agents as an especially valuable approach for engaging potential readers of the poster:~\textit{“It gives you good suggestions, and based on the suggestions, you can just pretty much click through everything and then make small adjustments (...) when using it for a fast-paced project, I think I could use it more conveniently.”}~(P10)

On the other hand, some participants highlighted \system{}'s potential as a prototyping tool, enabling designers to foresee potential challenges throughout the design process. They emphasized the value of the diverse perspectives offered by \system{}, noting its ability to identify and address issues early, before advancing to higher-fidelity stages: \textit{“I think it's a good strategy for me to obtain to kind of like, foresee the problem that might occur during the poster design process and then address them beforehand.”}~(P5)

However, they also pointed out the need for further iterations before the posters could be fully deployed. For instance, while they were generally positive about the ability to scale up the image generation that text-to-image models offer, they also noted that these images might lack realism and did not feature the specific product image or brand information. As such, they suggested that the generated images could serve as a reference for capturing the final image during the last iteration:~\textit{“It (using an AI-generated image) might be a good point to start or get the point, but I'd actually make it less AI-looking (...) it can be more like a reference or directional point, rather than exactly using it.”}~(P6)
\section{Controlled Online Evaluation}\label{sec:crowdsourcing}

Our user study demonstrated the effectiveness of \system{} within designers' workflows and revealed opportunities for future enhancements to better support their needs. To more precisely assess whether each agent (\ie{}, individual persona agent, moderator) fulfills its intended role in the design collaboration, we conducted a controlled evaluation. Specifically, we conducted two studies, each of which examined whether (1) the conclusion effectively synthesizes feedback from multiple audience personas, and (2) the feedback provided by each persona agent aligned appropriately with its designated persona identity.

\begin{figure*}
    \centering
    \includegraphics[width=.8\linewidth]{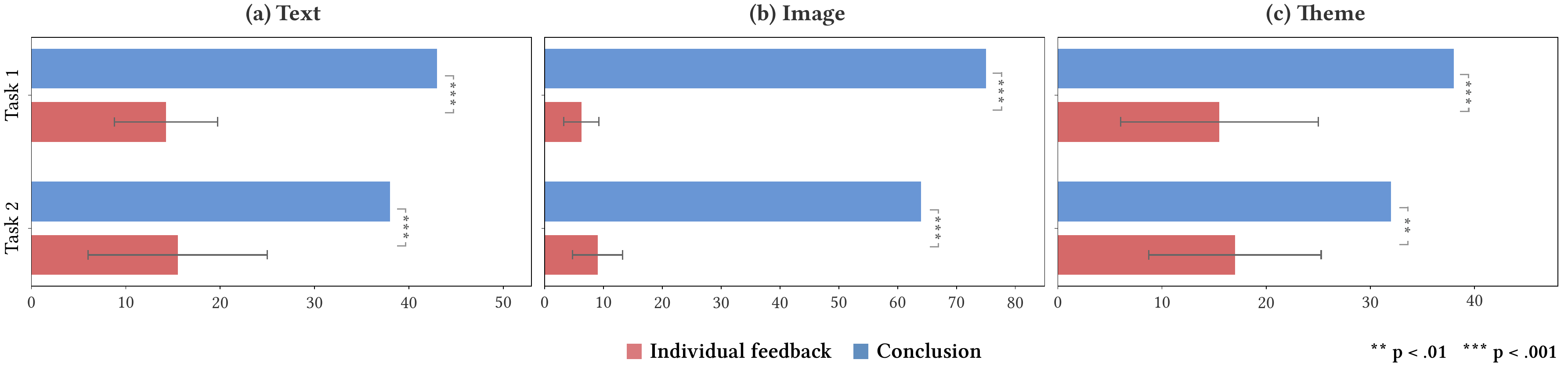}
    \caption{Evaluators' evaluation of the outcome from design feedback they believed would be most satisfying to the majority of target personas in our controlled experiment. For every data type, they perceived the conclusion achieved by \system{} to have best met the needs of the majority of personas. The tick labels indicate the count of the evaluators, and the gray bar indicates the standard deviation.}
    \label{fig:quant_crowdsourcing_consensus}
    \Description{Figure 9 contains three bar charts comparing evaluator preferences for individual feedback versus the moderated conclusion for (a) text, (b) image, and (c) theme. In all three charts, the bar representing the conclusion is significantly higher than the bar for individual feedback, indicating that evaluators found the conclusion derived from discussion to be more satisfactory for the majority of personas. Error bars show the standard deviation.}
\end{figure*}

\subsection{Methodology}

The studies were conducted online. Each evaluator went through both studies, each of which involved the evaluation of each data type that \system{} offers feedback on (\ie{}, text, image, and theme). To develop the questionnaires, we prepared the same initial poster drafts as we used in Section~\ref{sec:user_study}. Then, we provided these data to \system{} as input, where we extracted the returned feedback from four individual agents and a conclusion from their discussion to create questionnaires.

Each evaluator was then presented with the questionnaires, as well as the complete description of all persona agents, the original poster design, and the output of the feedback/conclusion. The detailed procedure for constructing the questionnaires and conducting a study for each study is as follows:

\subsubsection{Study 1: Assessing whether the conclusion effectively synthesizes the persona agents' feedback}

For each of the two poster designs, evaluators received three questionnaires, each focusing on each data type for feedback (\ie{}, text, image, and theme). Within each questionnaire, they were shown five outputs of the feedback for the corresponding data type---four derived from the individual feedback from each of the four persona agents, and one derived from the agents' discussion under the guidance of the moderator agent (without any user comment for consistency). The evaluators' task was to choose one of the feedback outputs most likely to satisfy the preferences of the majority of personas.

\subsubsection{Study 2: Assessing the appropriateness of feedback based on the agents' persona identity}

For each of the two poster designs, the evaluators were presented with the output of the feedback from each of the four persona agents across the three data types. Specifically, for each data type, they reviewed four outputs (each of which was from each persona agent) and were asked to identify which persona had provided the feedback leading to each output. The correct answer was the persona whose feedback corresponds to the presented output.

\subsection{Participants}

We recruited human evaluators ($N=100$) through Prolific\footnote{https://prolific.com}, an online human subjects pool. The average age of evaluators was 38.6 years ($SD=12.9$); 54 self-identified as female, 45 as male, and 1 preferred not to say. Since our study materials were in English, we required evaluators to be native English speakers based in the US. The median time spent for evaluation was 15.5 minutes, and evaluators were compensated an hourly wage of 11.3 USD.

\subsection{Results}

\subsubsection{Study 1: Effectiveness of synthesizing conclusion}

For all data types (\ie{}, text, image, and theme), evaluators found the conclusion from the discussion generated by \system{} to be most satisfactory to the majority of personas across both designs, in comparison to the individual feedback from each persona agent (see \autoref{fig:quant_crowdsourcing_consensus}).

For the text task, $40.5\%$ ($SD=2.5\%$) of evaluators selected the conclusion provided by \system{} as the most satisfactory across personas, compared to individual persona agent feedback ($M=14.9\%$, $SD=7.8\%$). A chi-square goodness-of-fit test confirmed that the distribution of responses significantly deviated (task 1: $\chi^2(4)=39.2$, $p<.001$; task 2: $\chi^2(4)=38.5$, $p<.001$). A post-hoc test further showed that the proportion of evaluators choosing \system{}'s conclusion was significantly higher than the combined alternatives (task 1: $z=4.65$, $p<.001$; task 2: $z=3.71$, $p<.001$), indicating that evaluators were more likely to find \system{}'s output most satisfactory across personas than the others (\autoref{fig:quant_crowdsourcing_consensus}a).

Similarly, for the image task, $69.5\%$ ($SD=5.5\%$) of evaluators selected the conclusion from \system{} as the most satisfactory across personas, compared to the individual feedback baseline ($M=7.6\%$, $SD=4.0\%$). The results indicated that the distribution of responses significantly deviated (task 1: $\chi^2(4)=191.0$, $p<.001$; task 2: $\chi^2(4)=124.8$, $p<.001$), and that the proportion of evaluators choosing \system{}'s output was significantly higher (task 1: $z = 12.7$, $p < 0.001$, task 2: $z = 9.17$, $p < 0.001$; \autoref{fig:quant_crowdsourcing_consensus}b).

Lastly, $42.5\%$ ($SD=10.5\%$) of evaluators rated the theme-based conclusion from \system{} as the most satisfactory for the majority of personas, compared to the individual feedback condition ($M=14.3\%$, $SD=8.2\%$). The results revealed significant deviations from uniform choice distributions (task 1: $\chi^2(4)=78.5$, $p<.001$; task 2: $\chi^2(4)=22.9$, $p<.001$), and a post-hoc test confirmed that \system{}'s conclusion was selected significantly more (task 1: $z = 6.61$, $p < 0.001$, task 2: $z = 2.57$, $p < 0.01$; \autoref{fig:quant_crowdsourcing_consensus}c).

\subsubsection{Study 2: Appropriateness of feedback based on the agents' persona identity}

\begin{figure*}
    \centering
    \includegraphics[width=.6\linewidth]{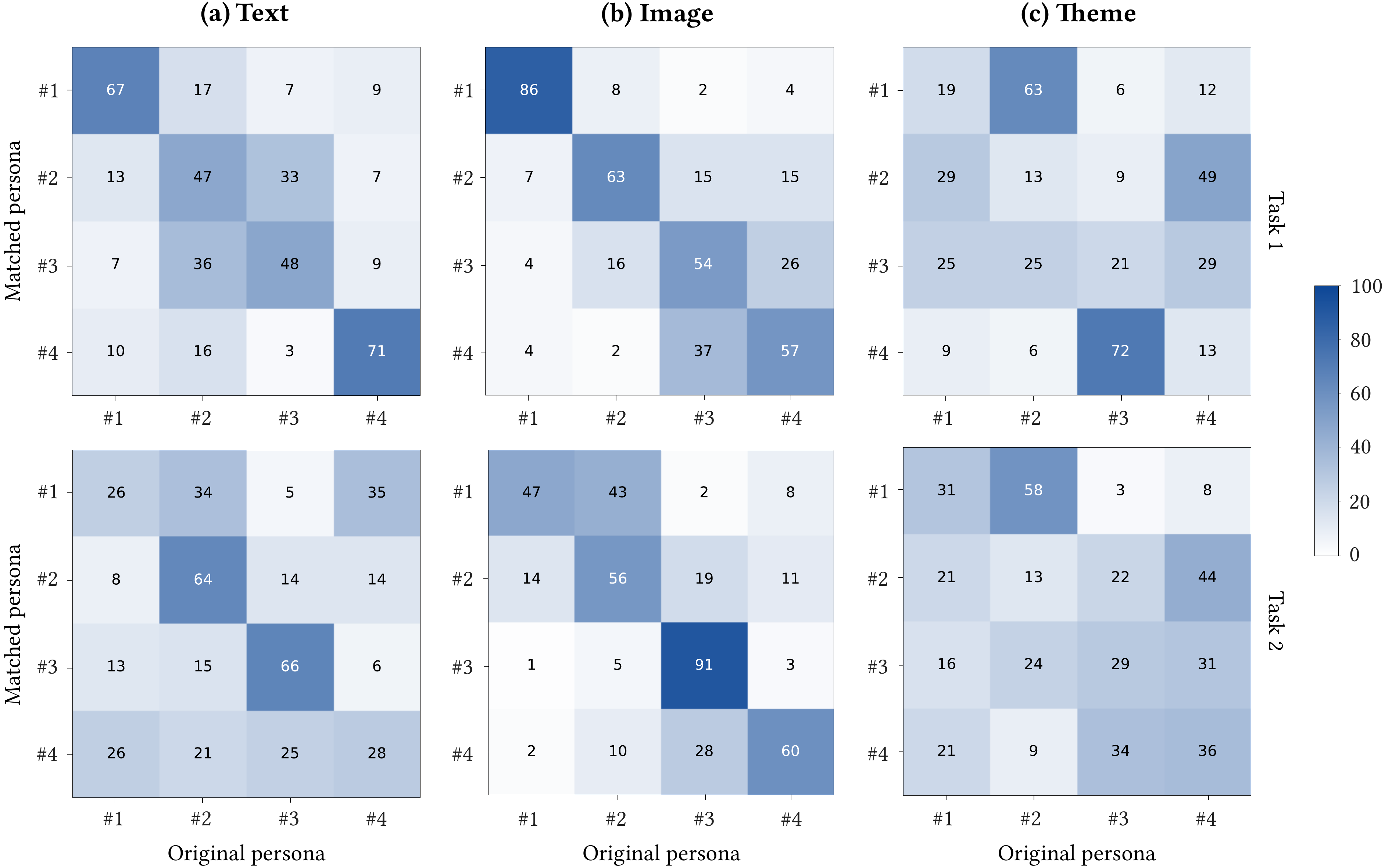}
    \caption{Evaluators' evaluations of the matching between persona descriptions and the source of each feedback's output from our controlled experiment. The x-axis represents the actual persona agent that made the feedback, while the y-axis indicates the persona agent the evaluator guessed would have made that feedback. The more concentrated the color along the diagonal (\ie{}, slanting downward from left to right), the more accurately evaluators matched the personas correctly. The number in each cell indicates the count of evaluators.}
    \label{fig:quant_crowdsourcing_match}
    \Description{Figure 10 presents a series of six confusion matrices from a controlled experiment, assessing evaluators' ability to match feedback to its source persona. The x-axis shows the actual persona agent who gave the feedback, and the y-axis shows the persona agent the evaluator guessed.}
\end{figure*}

In addition to assessing the synthesis of the conclusion, we could also identify that the feedback from the individual agents on the text and image appropriately reflected their identity (see \autoref{fig:quant_crowdsourcing_match}).

On average, $52.1\%$ of evaluators ($SD=16.7\%$) correctly matched the persona detail with each output from the provided feedback on the text. Here, if the evaluators had made their selection randomly, the expected hit rate would have been $25.0\%$. Similarly, $64.3\%$ ($SD=14.7\%$) of evaluators on average successfully identified the original persona for the outcomes from the image feedback. However, our findings suggest that evaluators encountered difficulties in accurately matching the provenance of the recommended themes, implying that a single theme may be perceived as satisfactory to multiple personas and may not be easily attributed to the characteristics of individual personas distinctly. On average, only $21.9\%$ of evaluators ($SD = 8.5\%$) were able to correctly match the presented themes with the persona that recommended them.
\section{Discussion}

Through our studies, we illustrated the promise of using audience-driven persona agents as design collaborators, as well as the efficacy of \system{}-supported designing in enhancing poster designs. By systematically supporting the construction of diverse persona agents that can engage in design discussions, users of \system{} reported being able to better understand aspects of the target audience and marketing details. Furthermore, by facilitating the discussion process and providing actionable insights, \system{} was viewed as an efficient and effective tool for reconciling perspectives from various audience groups. 

One of our findings from the user study was the importance of both enabling the efficient understanding of feedback with contextualized outputs (\eg{}, previews) and helping designers understand the rationale behind such feedback. Firstly, describing themselves as “visual learners,” participants from our study emphasized the efficacy of having previews to assist them in quickly comprehending feedback, as well as the need for further tailoring such visual supports to each data type of feedback. On the other hand, participants noted that the explainability of \system{}, facilitated through dialogues of the persona agents, enhanced their understanding of the motivations behind each persona agent's feedback, which was crucial for decision-making amid suggestions provided by a variety of persona agents. This implies that effective digestion of feedback and the qualitative understanding of its provenance are both essential for facilitating audience-driven design. With this in mind, we believe that the use of \system{} could be further improved by incorporating the previews of images and themes as an actual visual output, while presenting detailed provenance of feedback generated by persona agents associated with the persona details optionally presented through tooltips or an accordion panel to improve explainability without increasing cognitive load.

One question that arose from our study is how to better align the theme feedback from persona agents (\ie{}, tone, color) with the most relevant set of consolidated design themes. A survey from our user study indicated a challenge in accurately translating the theme feedback and discussion-driven conclusion, represented as tone and color, into the actual theme outputs. This issue also impacted the relationship between the characteristics of persona agents and the returned list of themes, where the evaluators in our controlled assessment had difficulty linking the theme outputs to their respective provenance agents. We presume this partly stems from our design choice to pull up from existing repositories. Although our system leveraged a large number of established themes to ensure a coherent style based on other designers' work, there remains a possibility of not achieving a perfect match in tone/color. We believe this issue can be mitigated by expanding the available themes by combining our existing sets with machine-generated theme generation (\eg{},~\cite{guo2021vinci}) and scaling up the set. Also, we observed a user in our study who quickly adjusted the background color of the applied theme to better match the result generated by \system{}, effectively bypassing the limitations of automated suggestions. This implies that proactively encouraging users to freely customize theme elements (\eg{}, colors, embellishments) in line with the feedback could mitigate this issue and help them achieve results that more closely match their expectations.

While \system{} relies on marketing briefs to generate the persona agents, our design allows it to leverage other sources of information if marketing briefs are not available or if other sources of information are available. For example, it is a common practice for designers to use diagramming tools (\eg{}, Miro~\cite{miro}, FigJam~\cite{figjam}) to structure their design process~\cite{feng2023understanding}. \system{} could use the information available in these other tools to infer design goals and target audiences. By connecting with these tools, we could substitute the input source while still constructing persona agents. Additionally, we could also envision supporting the manual entry of marketing details by providing guided text fields for key information (\eg{}, general target audience, constraints), which could later be used to construct persona agents.

\edited{Additionally, we designed \system{} with the goal of preserving user agency in mind, and future work on agent-supported design systems like \system{} could build on this by incorporating additional layers that further strengthen it. Prior works have identified design fixation and diminished agency as key concerns in AI-assisted creativity tools (\eg{},~\cite{wadinambiarachchi2024effects, knearem2023exploring}). With this in mind, we positioned \system{} as a collaborative assistant, promoting user control rather than automating decisions. We believe such considerations contributed to participants' selective engagement with persona feedback, such as iterating, editing, or disregarding suggestions at their own discretion (\autoref{fig:interaction_sequence}), maintaining their agency while avoiding overrelying on AI. Moving forward, future work incorporating audience-driven agents could further support this by introducing features to balance guidance with creativity, such as including a Devil's Advocate agent for challenging recommendations.}

Although \system{} is focused on supporting poster creation, we believe our general approach has the potential to extend into other design domains. One area where the concept of \system{} could be advantageous is in user interface (UI) design. Typically driven by specific user metrics (\eg{}, user engagement of target groups), UI design could similarly benefit from incorporating multiple persona agents that engage in discussions to ensure the design meets the needs of diverse target users. Similarly, by effectively structuring the design workspace as a data representation and interfacing with the agents, we believe future works could leverage our concept to benefit other goal-oriented design domains.

\section{Limitations and Future Work}

As outlined in our system description, we determined the number of dimensions for steering the set of persona agents based on the prior literature and computational scalability. Future work may investigate how changing the number of dimensions for constructing persona agents may potentially impact computational scalability and audience diversity.

In our controlled online experiment, we examined the effectiveness of discussions without providing user comments. This approach helped minimize noise from the variability of user comments, but it may not accurately represent real-world discussions among persona agents. Thus, future research could explore the alignment of results when user comments are incorporated.

\edited{Additionally, our online evaluation primarily focused on the process-oriented benefits of \system{} as an initial demonstration of the audience-driven persona agent concept. To more rigorously establish its effectiveness, future work should directly assess the final design artifacts (\ie{}, by comparing outputs generated with \system{} against those created without it or using alternative AI tools; \eg{}, designers working from the same brief without AI assistance, or designers using a single-shot LLM prompt for general poster improvements). Conducting such comparative studies would be a key area for future research to provide stronger empirical validation.}

\edited{Similarly, although participants recognized the distinct origins of feedback, engaged with its motivations, and appreciated the diversity of audience-driven responses, further causal validation is necessary to confirm that detailed persona attributes specifically shape the feedback. Future research could employ additional methods (\eg{}, ablation studies) to more deeply examine this relationship.}

Lastly, while we selected two marketing briefs for our study to facilitate participant comparisons, \edited{and although we found that \system{} consistently generated personas without including irrelevant dimensions, evaluating it with a wider range of typical and atypical materials would better validate the \system{}'s generalizability}. Similarly, \system{} takes a single brief as input which serves as a singular directive provided by marketers, and future works may investigate how to reconcile multiple briefs to form a new marketing basis and integrate into our concept.
\section{Conclusion}

In this study, we present \system{}, a collaborative assistant that leverages target audiences represented as persona agents from a marketing document as design collaborators. Harnessing generative AI, \system{} creates persona agents based on a marketing document to construct diverse perspectives. With these persona agents, \system{} is designed to help users obtain diverse audience design feedback rooted in the marketing document, effectively reconcile diverse perspectives, and enrich the poster design process. Our exploratory user study and controlled study demonstrated that the agents in \system{} could successfully play the expected role of design collaborators, as well as the system's efficacy in helping users to notice viewpoints that might have been overlooked.

\bibliographystyle{ACM-Reference-Format}
\bibliography{100-bibliography}

\newpage
\appendix
\section{Study Details}

\subsection{Formative Interview Study}
\begin{table}[h!]
    \caption{Participants in our formative interview study}
    \footnotesize
    \centering
    \begin{tabular}{llll}
    \toprule
    \textbf{PID} & \textbf{Gender} & \textbf{Age} & \textbf{\leftcell{Design\\experience}} \\
    \midrule
    FP1 & Female & 25-34 & 2 years \\
    FP2 & Female & 25-34 & 5 years \\
    FP3 & Male & 18-24 & 2 years \\
    FP4 & Female & 25-34 & 9 years \\
    FP5 & Female & 25-34 & 5 years \\
    FP6 & Female & 18-24 & 5 years \\
    FP7 & Female & 18-24 & 2 years\\
    FP8 & Male & 18-24 & 2 years \\
    \bottomrule
    \end{tabular}
    \label{tab:participants}
    \Description{Table 2 lists the demographic information of the eight participants (FP1-FP8) in the formative interview study. The table has four columns: participant ID (PID), gender, age range, and years of design experience.}
\end{table}

\subsection{User Study \& Controlled Online Evaluation}\label{sec:study_details}

\begin{figure}[h!]
  \centering
  \begin{subfigure}{0.43\linewidth}
    \centering
    \fbox{\includegraphics[width=\linewidth]{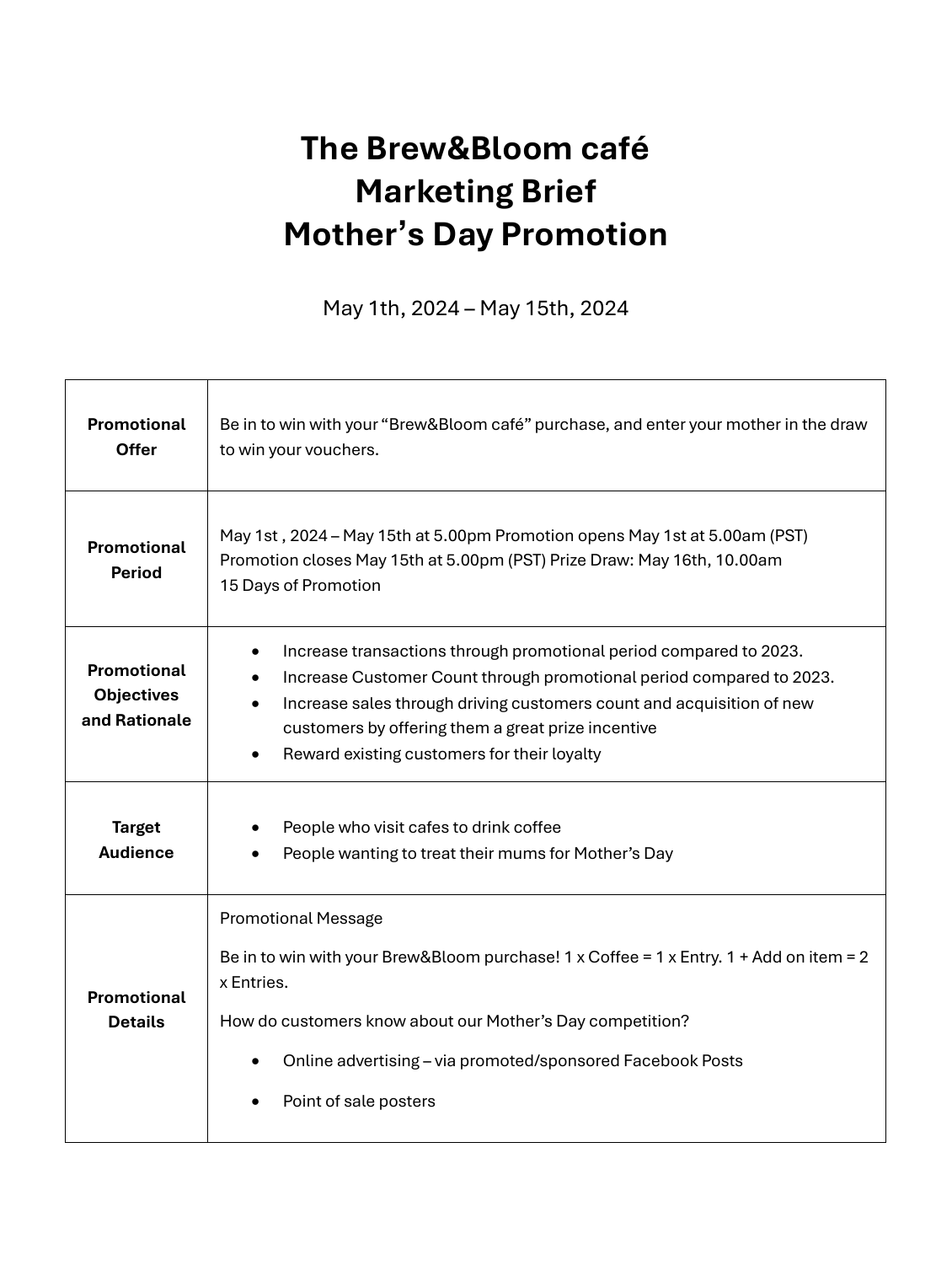}}
    \caption{Marketing brief 1}
    \label{fig:1}
  \end{subfigure}
  \hfill 
  \begin{subfigure}{0.43\linewidth}
    \centering
    \fbox{\includegraphics[width=\linewidth]{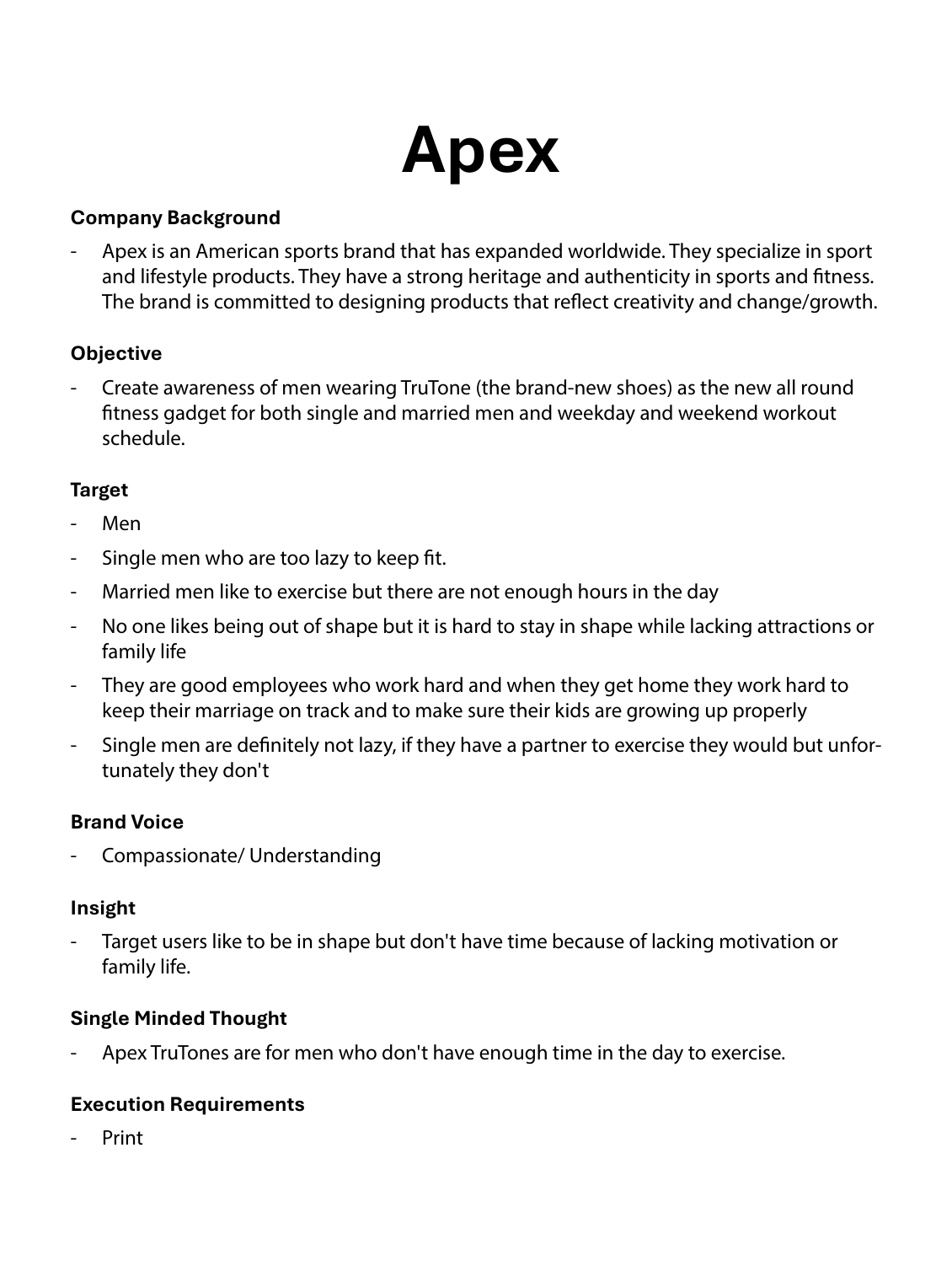}}
    \caption{Marketing brief 2}
    \label{fig:2}
  \end{subfigure}
\caption{Marketing briefs used in our studies}
\label{fig:brief}
\Description{Figure 11 shows the two marketing briefs used in the user studies. Figure 11a is the brief for the café. Figure 11b is the brief for a sports brand.}
\end{figure}

\begin{figure}[h!]
  \centering
  \begin{subfigure}{0.43\linewidth}
    \centering
    \fbox{\includegraphics[width=\linewidth]{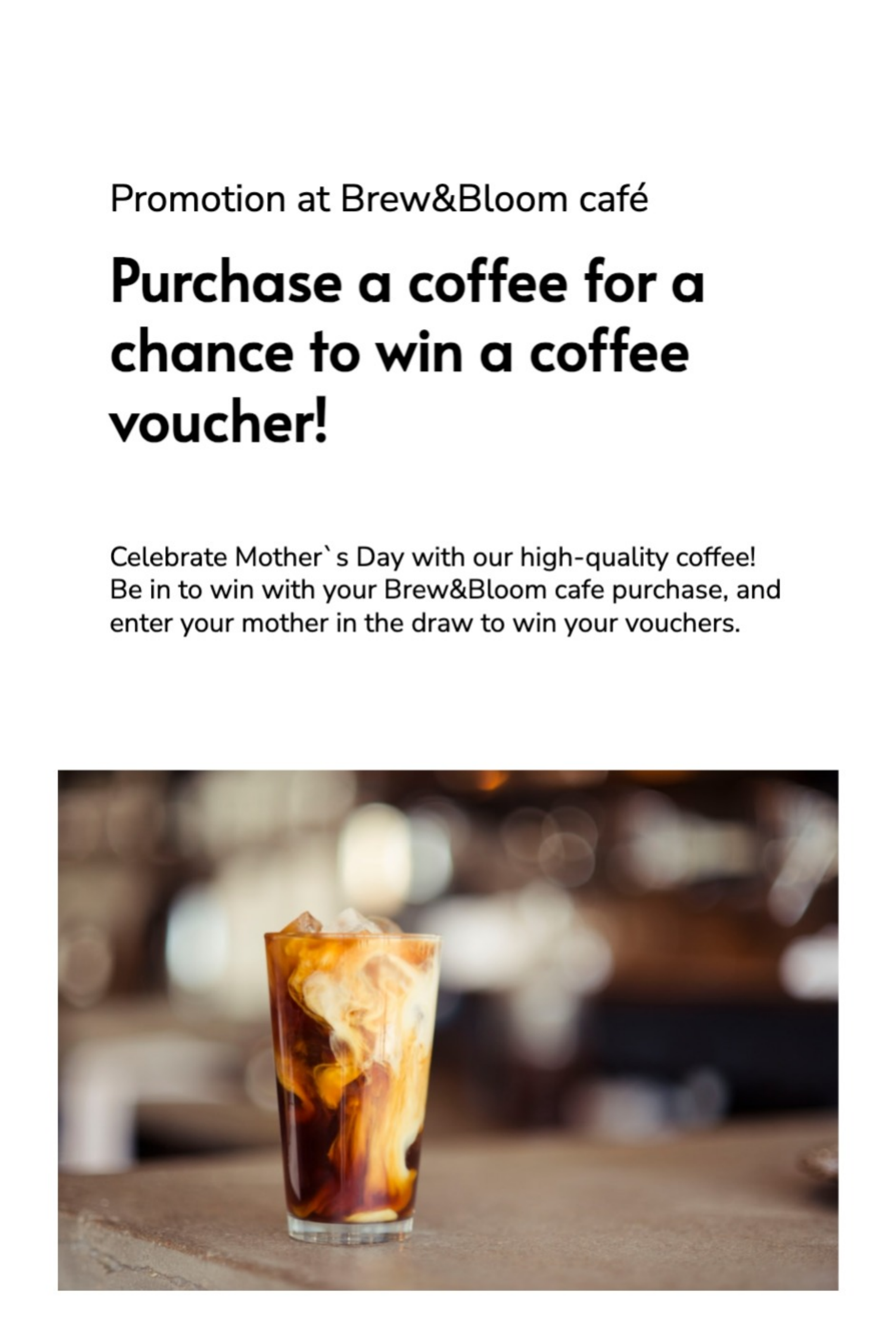}}
    \caption{Poster draft used for marketing brief 1}
    \label{fig:1}
  \end{subfigure}
  \hfill 
  \begin{subfigure}{0.43\linewidth}
    \centering
    \fbox{\includegraphics[width=\linewidth]{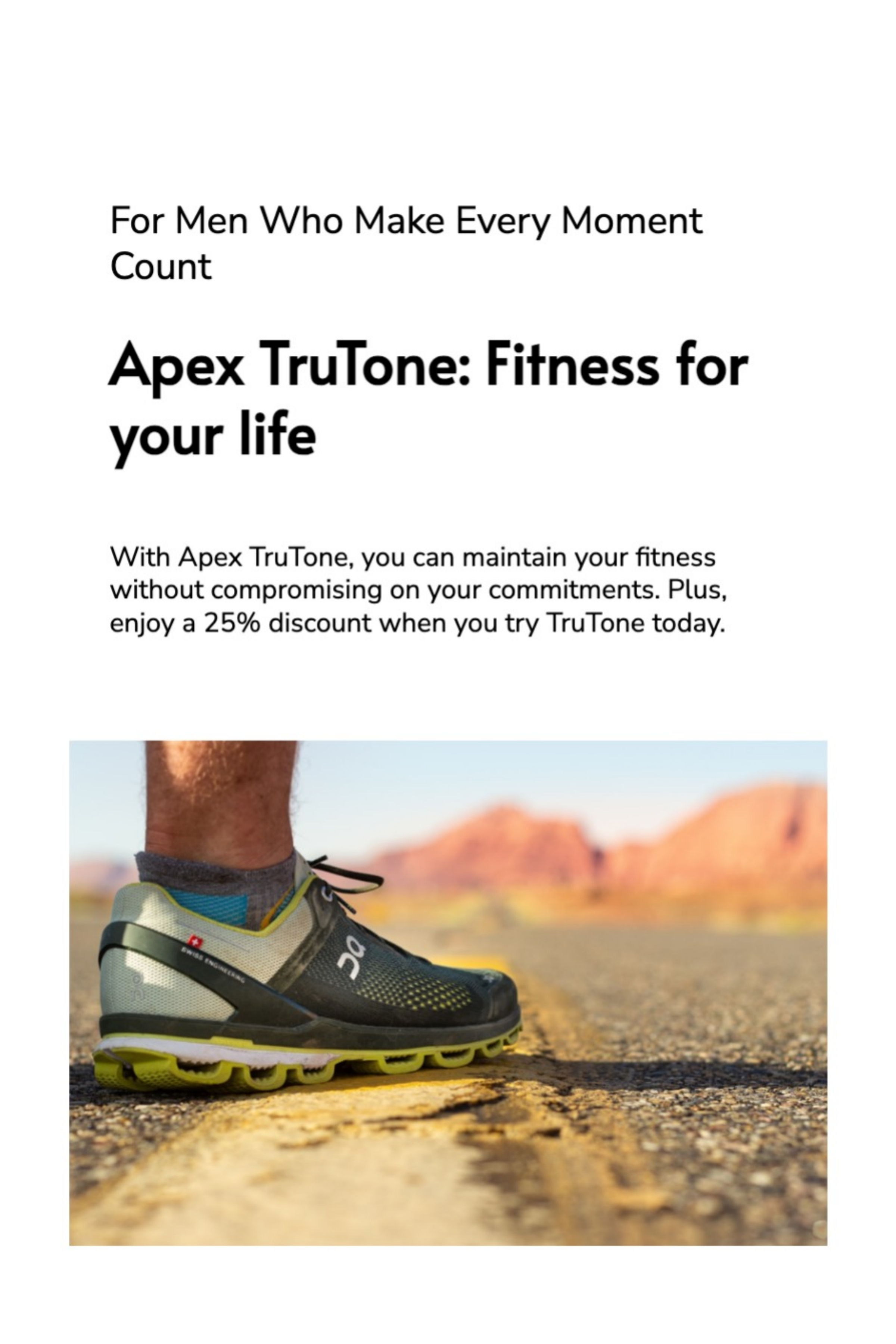}}
    \caption{Poster draft used for marketing brief 2}
    \label{fig:2}
  \end{subfigure}
\caption{\edited{Poster drafts used in our studies}}
\label{fig:draft}
\Description{Figure 12 displays the two initial poster drafts that were provided to study participants. Figure 12a is the draft for the café brief, featuring simple text and a coffee image. Figure 12b is the draft for the sports brand brief, with basic text and an image of a man exercising.}
\end{figure}

\begin{figure*}
    \centering
    \includegraphics[width=.8\textwidth]{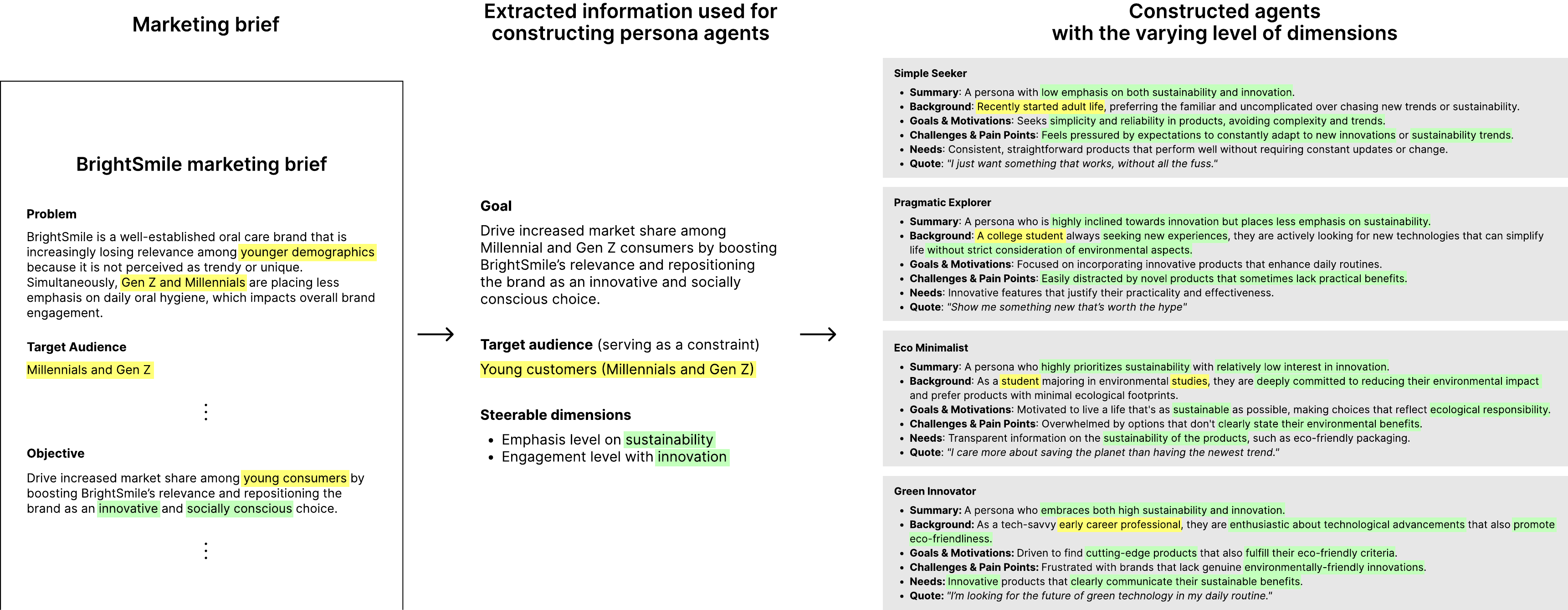}
    \caption{An example of a marketing brief and how persona agents could be created from the marketing brief. The contents could be extracted to construct target audiences as interactive agents with unique perspectives by building on the shared foundation (\ie{}, target audience description) with steerable dimensions. For the dimensions, \system{} prioritizes dimensions present in the brief, while generating contextually relevant ones if none exist; in both cases, \system{} is prompted to steer dimensions whose extremes do not contradict marketing brief details.}
    \label{fig:marketing-brief}
    \Description{Figure 13 is a workflow diagram explaining how persona agents are created from a marketing brief. On the left, a marketing brief has key information highlighted. In the middle, this information is shown as extracted information, including the goal, target audience, and two steerable dimensions. On the right, this information is used to construct four detailed persona agent profiles, each representing a unique combination of the steerable dimensions.}
\end{figure*}

\section{Example Discussion}
\label{sec:example_discussion}
We provide an example discussion that \system{} simulated based on the marketing brief for a café (\autoref{fig:brief}b), with only two agents involved for illustrative purposes. The conflict between two persona agents is about “\textit{Conflicting views on the level of emphasis on ‘Mother's Day’ in the promotional text,}” as the persona agent \#1's feedback was “\textit{Subtly reference ‘Mother's Day’ in the text, which might resonate with customers looking for a reason to treat their mothers without being overly promotional.}” and the persona agent \#2's feedback was “\textit{Changing the text to incorporate the special occasion of Mother's Day explicitly.}”

The conversation below highlights the \textcolor{cyan}{thought-provoking questions} and \textcolor{orange}{how the conclusion is driven} from the agents' responses.

\begin{itemize}
    \item \textbf{Moderator Agent}: \textcolor{cyan}{How might a more subtle mention of ‘Mother's Day' help in attracting new customers as opposed to a more explicit mention?} \\
    \textbf{Persona Agent \#1}: \textit{A more subtle mention of Mother's Day can keep the promotion aligned with the café’s peaceful and relaxing atmosphere, appealing to regular customers like me who prefer a low-key experience. For example, the text could read, ‘Enjoy a special coffee break this May, and you might treat your mother with a surprise gift!’ This still honors the theme without overwhelming customers with heavy promotions.}
    \item \textbf{Moderator Agent}: \textcolor{cyan}{How do you think emphasizing Mother’s Day explicitly in the promotional text would influence the engagement of existing customers versus attracting new ones?} \\
    \textbf{Persona Agent \#2}: \textit{Explicitly mentioning Mother's Day in the promotional text can make the offer more attractive to occasional visitors like myself. It clearly communicates the special occasion and makes it immediately relevant, encouraging people who may not regularly visit the café to make a special trip. Reflecting this, the promotional message could be modified to something like ‘Celebrate Mother’s Day with Brew\&Bloom! Purchase a coffee and enter your mum into the draw to win a coffee voucher!' to clearly convey the incentive without overwhelming frequent visitors.}
    \item \textbf{Moderator Agent}: \textcolor{orange}{Based on the given responses, I think the promotional text should mention Mother's Day explicitly but in a way that balances appealing to both new and regular customers without overwhelming them.}
\end{itemize}

The resulting summary of the conclusion would be \textit{“Modifying the promotional text to emphasize Mother's Day, while ensuring that it remains inviting and not overly promotional.”}

\end{document}